\documentclass[twoside,11pt]{article}

\usepackage{jmlr2e}

\usepackage{graphicx}
\usepackage{amsthm}
\usepackage{amssymb}
\usepackage{mathrsfs}
\usepackage{setspace}
\usepackage{amsmath}
\usepackage{commath}
\usepackage{xfrac}
\usepackage{booktabs}
\usepackage{siunitx}
\usepackage{rotating}
\usepackage{tikz} 
\usepackage{etoolbox}
\usepackage{bm}
\usepackage[utf8]{inputenc} 
\usepackage{float}
\usepackage{epsfig}
\usepackage{hyperref}
\usepackage[nameinlink,noabbrev]{cleveref}

\input xy
\xyoption{all}

\hypersetup{%
  colorlinks=true,
  linkcolor=blue,
  linkbordercolor=red,
  citecolor=blue,
  linkcolor = blue,
  anchorcolor = blue,
  filecolor = blue,
  urlcolor = red
}

\makeatletter
\renewcommand\paragraph{\@startsection{paragraph}{4}{\z@}%
            {-2.5ex\@plus -1ex \@minus -.25ex}%
            {1.25ex \@plus .25ex}%
            {\scshape\small\bfseries}}
\makeatother
\jmlrheading{1}{2020}{1-48}{1/20}{10/00}{Rahul Radhakrishnan Iyer, Manish Sharma and Vijaya Saradhi}

\ShortHeadings{Conference Recommender System for Authors}{Iyer, Sharma and Saradhi}
\firstpageno{1}

\begin{document}

\title{A Correspondence Analysis Framework for Author-Conference Recommendations}
\author{\name Rahul Radhakrishnan Iyer\thanks{Corresponding Author} \email rahuli@alumni.cmu.edu \\
       \addr Language Technologies Institute\\
       Carnegie Mellon University\\
       Pittsburgh, PA 15213, USA
       \AND
       \name Manish Sharma \email mssharma5523@gmail.com \\
       \addr Microsoft Corporation India Pvt. Ltd.\\
       Banjara Hills, \\
       Hyderabad, Telangana 500034, India
       \AND
       \name Vijaya Saradhi \email saradhi@iitg.ac.in \\
       \addr Department of Computer Science and Engineering\\
       Indian Institute of Technology Guwahati\\
       Guwahati, Assam 781039, India}


\maketitle

\begin{abstract}
For many years, achievements and discoveries made by scientists are made aware through research papers published in appropriate journals or conferences. Often, established scientists and especially newbies are caught up in the dilemma of choosing an appropriate conference to get their work through. Every scientific conference and journal is inclined towards a particular field of research and there is a vast multitude of them for any particular field. Choosing an appropriate venue is vital as it helps in reaching out to the right audience and also to further one's chance of getting their paper published. In this work, we address the problem of recommending appropriate conferences to the authors to increase their chances of acceptance. We present three different approaches for the same involving the use of social network of the authors and the content of the paper in the settings of dimensionality reduction and topic modeling. In all these approaches, we apply Correspondence Analysis (CA) to derive appropriate relationships between the entities in question, such as conferences and papers. Our models show promising results when compared with existing methods such as content-based filtering, collaborative filtering and hybrid filtering.

\end{abstract}


\section{Introduction}
\label{sec:introduction}
With the advent of the Internet and the growing amount of information available therein, people are increasingly resorting to finding information online. This in turn has resulted in several challenges, one of the main ones for users being finding exactly what they are looking for or for researchers to keep upto date on information of whose existence they may be unaware.\\

For many years, achievements and discoveries made by scientists are made aware through research papers published in appropriate journals or conferences. Often, established scientists and especially newbies are caught up in the dilemma of choosing an appropriate conference to get their work through. Every scientific conference and journal is inclined towards a particular field of research and there is a vast multitude of them for any particular field. Choosing an appropriate venue is vital as it helps in reaching out to the right audience and also to further one's chance of getting their paper published. \\

In order to address this problem, we aim to build a recommender system that recommends the most appropriate publication venues for an author. This system is particularly useful to budding researchers who have very little knowledge about the research world and also to experienced researchers by saving a lot of their time and effort.  \\

In this work, we aim to approach this problem in the settings of dimensionality reduction and topic modeling. We propose three different methods to recommend conferences for researchers to submit their paper based on the content of the paper and the social network of the authors: two of them involving content-analysis and the third one involving social network of the authors. Our approach is empirically evaluated using a dataset of recent ACM conference publications and compared with existing methods such as content-based filtering, collaborative filtering and hybrid filtering with promising results.\\

However, there are several challenges that need to be addressed. We list out the challenges along with the different claims from our work\\

\begin{enumerate}
	\item \textbf{Challenges:} We face several challenges when working in this domain, as illustrated.
		\begin{enumerate}
			\item In recent times, using dimensionality reduction methods such as SVD, PCA are becoming widely popular in application to recommender systems. The use of Correspondence Analysis (CA) has not been explored as much in the literature. In some of the recent works, PCA, which can only be applied to continuous data, has been applied to tabulated discrete data. How do we remedy this defect? 
			\item In all the previous work done related to our problem, only a model using the social network of the authors has been employed. Content analysis of the papers in consideration, to the best of the authors' knowledge, has never been attempted. Just using the network of authors, without even looking at the paper, is not sufficient to decide where the paper should go to. How do we incorporate content into our work?
			\item Suggesting conferences to new authors is a very tricky business. If the author has not published any paper before, he does not have a social network. Hence, the current systems would yield a poor recommendation. Will considering content of the paper lead to better results?
			
			\item Constructing matrices in higher dimensional spaces, as in our case, invites a large amount of redundancy and hence, the  relationship between the two attributes in consideration is not obtained with clarity. How can this problem be tackled?
			
			\item For the second method in our work, we construct a Paper $\times$ Words matrix and a Words $\times$ Conference matrix, where the $\left(i, j\right)^{th}$ entry of each of the matrices indicate the frequency of occurrence of $word_j$ in $paper_i$ and $word_i$ in the papers published in $conference_j$ respectively. For the process of recommendation, we compose the two matrices to obtain a Paper $\times$ Conference on which we apply CA to proceed. But it is not guaranteed that the entries in the matrix obtained are the frequencies. How can we make sure this question does not arise? 
		\end{enumerate}
	
	\item \textbf{Main Claims:} We use the abstracts of the papers in consideration for content analysis. The challenges raised above are systematically addressed as follows:
		\begin{enumerate}
			\item In our work, we deal with tabulated discrete data. From all the data collected, we construct matrices such as Paper $\times$ Words, Paper $\times$ Conference, Words $\times$ Conference. Each entry of these matrices represents the frequency in question and thus forms the basis of our methods in applying dimensionality reduction techniques. We remedy the continuous data conundrum with the use of Correspondence Analysis (CA). By reducing the matrices to lower dimensional subspaces using CA, we obtain the necessary relationship between the two entities with clarity, thus avoiding having to use PCA. This makes more the approach taken all the more meaningful.
			
			\item As suggested, just relying on the network of the authors is not sufficient to obtain a good recommendation of a conference. We bring in the content of the paper into our work to build a better model, which to the best of the authors' knowledge has not been explored before in the literature. Since the essence of the entire paper is contained within it's abstract, we build the content matrices using just the abstracts of the various papers. We employ term \textit{frequency-inverse document frequency} (tf-idf) to generate the matrices of important keywords from the abstracts. In two of the methods, we construct Paper $\times$ Words and Words $\times$ Conference matrices using the above mentioned technique.
			
			\item It would be problematic as suggested to recommend conferences to authors with no prior social network. But this problem does not arise during content-analysis as we are not concerned with the author's social network. Just relying on the content of the abstract, we recommend suitable conferences. In our experiments, to suggest conferences to new authors, we observed that this method far supersedes the one relying on only his/her social network.
			
			\item Maximum essence of the relationship between the attributes in a table is obtained only in lower dimensional subspaces. Thus, when reducing the dimension of the matrices using CA, we essentially throw out the redundant information while maintaining the crucial and important part of them that are responsible for the relationships. As an added bonus, the reduced dimension increases the efficiency of the methods.
			
			\item In order to avoid such a confusion, our third method does not compose the two matrices. Instead a linear transformation is defined between the two spaces after reduction of dimension. In essence, after constructing the Paper $\times$ Words and Words $\times$ Conference matrices, we apply CA to each of them to reduce their dimension and then define a linear transformation from one subspace to the other for the process of recommendation.
			
		\end{enumerate}
		
	\item \textbf{Key tasks of the methods:} The key tasks of each of the method proposed are listed as follows:
		\begin{enumerate}
			\item \textbf{Method 1:} Involving the use of the social network of the authors.
				\begin{itemize}
					\item We construct the Author $\times$ Conference matrix, with each row consisting of entries for a particular author and the $\left(i, j\right)^{th}$ entry of the matrix representing the number of times author $a_i$ has published in conference $c_j$.
					\item We apply CA on this matrix to obtain principal column co-ordinates corresponding to the conferences. Using this, we obtain the principal row co-ordinates corresponding to the authors, whose paper needs a conference recommendation.
					\item The conference nearest to the obtained author cluster in the bi-plot is recommended as the most suitable conference.
				\end{itemize}
			\item \textbf{Method 2:} Considering the content of the paper and composition of matrices.
				\begin{itemize}
					\item We construct a Paper $\times$ Words matrix and a Words $\times$ Conference matrix, where the $\left(i, j\right)^{th}$ entry of each of the matrices indicate the frequency of occurrence of $word_j$ in $paper_i$ and $word_i$ in the papers published in $conference_j$ respectively.
					\item Then, we compose these two matrices and apply CA to obtain the principal column co-ordinates corresponding to the conferences.
					\item We obtain the principal row co-ordinates of the paper in need of a recommendation by computing it's tf-idf vector, composing with the Words $\times$ Conference training matrix and subsequent CA.
					\item The conference nearest to the paper in the bi-plot is recommended as the most suitable one.
				\end{itemize}
			\item \textbf{Method 3:} Considering the content of the paper and a linear transformation.
				\begin{itemize}
					\item We construct the Paper $\times$ Words and Words $\times$ Conference matrices as before, but instead of composing them, we reduce them to lower dimensional subspaces individually using CA.
					\item Then, we define a linear transformation from the reduced paper space to the reduced conference space.
					\item This linear transformation enables us to take a paper, in need of recommendation, to the space of conferences and suggest a conference closest to it.
				\end{itemize}
		\end{enumerate}
\end{enumerate}

\subsection{Organization of the Paper}
\label{subsec:organization}
The paper is organized as follows. Section \ref{sec:related_works} gives an overview of the works related to the problem at hand. We formulate the problem and build on the main techniques used in the experiments in Sections \ref{sec:problem_statement} and \ref{sec:preliminary_concepts} respectively. Section \ref{sec:data_sets_tools} details the datasets and tools used. The technical approaches used and the experimental results obtained with their implications are discussed in Sections \ref{sec:work_done} and \ref{sec:eval} respectively. We finally present and draw conclusions with remarks, exploring possibilities and scopes of future work in Section \ref{sec:conclusions_future}.

\section{Related Works}
\label{sec:related_works}
The field of recommender systems, being recent, has been a hot topic for researchers in the last few years. A lot of work has been done in exploring different algorithms and techniques to aid in building systems that can make intelligent suggestions to consumers. There has been work in content-based filtering, collaborative filtering, knowledge-based recommender systems and data mining areas such as classification, clustering, association rule mining and dimensionality reduction.

\subsection{Collaborative Filtering}
\label{subsec:related_collaborative_filtering}
There have been many collaborative systems developed in the academia and the industry. Algorithms for collaborative recommendations can be grouped into two general classes: memory-based(or heuristic-based) and model-based. Memory-based algorithms essentially are heuristics that make rating predictions based on the entire collection of previously rated items by the users. That is, the value of the unknown rating $r_{c,s}$ for user $c$ and item $s$ is usually computed as an aggregate of the ratings of some other (usually, the $N$ most similar) users for the same item $s$.\\

There have been several model-based collaborative recommendation approaches proposed in the literature. These include a collaborative filtering method in a machine learning framework, where various machine learning techniques (such as artificial neural networks) coupled with feature extraction techniques (such as singular value decomposition — an algebraic technique for reducing dimensionality of matrices) are used. There have also been statistical models like Bayesian model and several algorithms for estimating parameters like $K$-means clustering and Gibbs sampling. More recently, a significant amount of research has been done in trying to model the recommendation process using more complex probabilistic models. Some probabilistic modeling techniques for recommender systems include Markov decision processes, probabilistic latent semantic analysis and a combination of multinomial mixture and aspect models using generative semantics of Latent Dirichlet Allocation.\\

Among the latest developments, techniques have been proposed to combine model-based and memory-based approach using probabilistic approaches. For example, 1) using an active learning approach to learn the probabilistic model of each user’s preferences and 2) using the stored user profiles in a mixture model to calculate recommendations.

\subsection{Content-based Filtering}
\label{subsec:content-based_filtering}
Content-based systems are designed mostly to recommend text-based items and the content in these systems is usually described with keywords. For example, a content-based component of the Fab \cite{balabanovic1997fab} system, which recommends Web pages to users, represents Web page content with the $100$ most important words. Similarly, the Syskill \& Webert system \cite{pazzani1997learning} represents documents with the $128$ most informative words. The “importance” (or “informativeness”) of word $k_j$ in document $d_j$ is determined with some weighting measure $w_ij$ that can be defined in several different ways.\\

\subsection{Dimensionality Reduction}
\label{subsec:dimensionality}
It is very common to see recommender systems use data with many features i.e. a very high-dimensional space. Despite the provision for many features, frequently it is observed that sparsity of the feature vectors is a common problem. This has many implications when it comes to clustering and outlier detection, where the notions of density and distance between points become less meaningful. This is often called \textit{the Curse of Dimensionality}. Dimensionality reduction techniques, thus, play an important role in these cases by helping to transform the original vectors into those in lower dimensional subspaces.\\

Xun Zhou et. al \cite{zhou2012personalized} propose a scalable algorithm for recommendations called Incremental ApproSVD, which combines the incremental SVD algorithm with the ApproSVD (Approximating the  SVD) proposed in one of their earlier works where they use ApproSVD to generate  personalized recommendations. This has been shown to outperform the traditional incremental SVD algorithm, when run on the MovieLens and Flixster dataset.\\

As the Netflix Prize competition demonstrated, matrix factorization models are superior to classic nearest-neighbour techniques for producing product recommendations. Yehuda Koren \cite{bell2007bellkor} proposed the BellKor Solution to the Netflix Grand Prize. The baseline predictors were improved, an extension of the neighbourhood model that addresses temporal dynamics was introduced, a new Restricted Boltzmann Machine (RBM) model was used with superior accuracy by conditioning the visible units and finally a new blending algorithm which is based on Gradient Boosted Decision Trees (GBDT) was introduced. Gabor Takacs and Istvan Pilaszy \cite{takacs2008unified} propose a hybrid approach that combines an improved Matrix Factorization (MF) method with the \textit{NSVD1} approach, familiarized by Paterek \cite{paterek2007improving}, resulting in a very accurate factor model. Further, they propose a unification  of the factor models and neighbourhood-based approaches, which improves the performance. Having run their method on the Netflix Prize dataset, they provide a very low RMSE, which outperforms all published single methods in the literature.\\

Osman and Ismail \cite{osmanli2011using} used tag similarity techniques in SVD-based recommender systems. To  improve the recommendation quality, content information of the items in the form of user given tags are used. To adopt tags to the normal SVD algorithm, they have reduced the three-dimensional matrix <user, item, tag> to three two-dimensional matrices: <user, item>,  <user, tag> and <item, tag>. These matrices are used to perform the SVM recommendation. This has shown to increase the performance.\\

Goldberg et al. \cite{goldberg2001eigentaste} proposed an approach to use PCA in the context of an online joke recommendation system. Their system, known as Eigentaste, starts from a standard matrix of user ratings to items. They then select their gauge set by choosing the subset of items for which all users had a rating. This new matrix is then used to compute the global correlation matrix where a standard 2-dimensional PCA is applied. Manolis and Konstantinos \cite{vozalis2007using} \cite{vozalis2007recommender} propose an algorithm called PCA-Demog, which applies PCA for Demographically enhanced prediction generation. The filtering algorithm proposed  applies PCA on user ratings and demographic data. Along with the algorithm, possible ways of combining it with different sources of filtering data is also discussed. They also describe in one of their other works, application of SVD on Item-based Collaborative Filtering. They describe two algorithms: The first algorithm uses SVD in order to reduce the dimension of the active item's neighbourhood. The second algorithm initially enhances Item -based Filtering with demographic information and then applies SVD at various points of the filtering procedure.\\

Sarwar et al. \cite{sarwar2000application} describe two different ways to use SVD in the context of RS. First, SVD can be used to uncover latent relations between customers and products. In order to accomplish this goal, they first fill the zeros in the user-item matrix with the item average rating and then normalize by subtracting the user average. This matrix is then factored using SVD and the resulting decomposition can be used after some trivial operations directly to compute the predictions. The other approach is to use the low-dimensional space resulting from the SVD to improve neighborhood formation for later use in a $k$NN approach. As described by Sarwar et al. \cite{sarwar2002incremental} in one of their other works, one of the big advantages of SVD is that there are incremental algorithms to compute an approximated decomposition. This allows to accept new users or ratings without having to recompute the model that had been built from previously existing data. The same idea was later extended and formalized by  Brand into an online SVD model, where he used these methods to model data streams describing tables of consumer/product ratings, where fragments of rows and columns  arrive in random order and individual table entries are arbitrarily added, revised, or retracted at any time.\\

Several hybrid approaches of Collaborative Filtering (CF) and Content-based Filtering (CB)  have been proposed to increase the accuracy of recommendations. Major drawback in these were that the two techniques were most often executed independently. Panagiotis Symeonidis \cite{symeonidis2008content}  proposed a Content-based Dimensionality Reduction method for recommendations, wherein a feature profile for a user is constructed based on both collaborative and content features. Latent Semantic Indexing (LSI) is then applied to reveal the dominant features of the user.  Recommendations are then provided according to this dimensionally-reduced feature profile. This method has shown to outperform well-known CF, CB and hybrid approaches. \\

Zanker et al. \cite{zanker2008evaluating} evaluate the use of different recommender systems for the purpose of tourism. They have considered different methods: Correspondence Analysis, Click-stream Sequence Analysis and Contingency Analysis.\\

There have been advances in text-based recommender systems too as described:\\

Scientists depend on literature search to find prior work that is relevant to their research ideas. In this context, Steven Bethard and Dan Jurafsky \cite{bethard2010should} introduce a retrieval model for literature search that incorporates a wide variety of factors important to researchers, and learns the weights of each of these factors by observing citation patterns. They introduce features like topical similarity and author behavioral patterns, and combine these with features from related work like citation count and recency of publication. They present an iterative process for learning weights for these features that alternates between retrieving articles with the current retrieval model, and updating model weights by training a supervised classifier on these articles. In a similar context, Lee Giles et. al \cite{giles1998citeseer} built CiteSeer, an autonomous citation indexing system, which indexes academic literature in electronic format. Published research papers on the World Wide Web, increasing in quantity daily, are often poorly organized and often exist in non-text forms (eg. Postscript). Due to this, significant amounts of time and effort are commonly needed to find interesting and relevant publications on the Web. CiteSeer, being a Web based information agent, helps alleviate this problem by assisting the user in the process of performing a scientific literature search. \\

Concept-based document recommendation for CiteSeer authors is explored by Kannan Chandrasekaran et. al \cite{chandrasekaran2008concept}. They present a novel way of representing the user profiles as trees of concepts and an algorithm for computing the similarity between the user profiles and document profiles using a tree-edit distance measure. This has shown to outperform a traditional vector-space model. Another way of recommending documents is using the implicit social network of researchers, as proposed by Cheng Chen et. al \cite{chen2008implicit}.\\

Bela Gipp and Jordan Beel \cite{gipp2009citation}, propose an approach for identifying similar documents that can be used to assist scientists in finding related work. The approach called Citation Proximity Analysis (CPA) is a further development of co-citation analysis, but in addition, considers the proximity of citations to each other within an article's full-text. The underlying idea is that the closer citations are to each other, the more likely it is that they are related. The CPA based approach has been shown to have higher precision with possibility of identifying related sections within documents, compared to existing approaches like bibliographic coupling, co-citation analysis or keyword based approaches. \\

Qi He et. al \cite{he2011citation} propose an approach for automatic recommendation of citations for a manuscript without author supervision. This reduces user burden, as the input to the system is just a query manuscript (without a bibliography), and the system automatically finds locations where citations are needed. They have shown the effectiveness of their approach with an extensive empirical evaluation using the CiteSeerX data set. They further propose in one of their other works, a context-aware citation recommender system \cite{he2010context}, which helps in citing good candidates at different local contexts in the paper. \\

Ming Zhang et. al \cite{zhang2008paper} present a recommender for scientific literatures based on semantic concept similarity computed from the collaborative tags. User profiles and item profiles are presented by these semantic concepts, and neighbour users are selected using collaborative filtering. Then, content-based filtering approach is used to  generate recommendation list from the papers these neighbour users tagged. Onur et. al \cite{kuccuktuncc2012direction} also address a similar problem of recommending papers on academic networks, but here they use a direction-aware (in the sense that they can be tuned to find either recent or traditional papers) citation analysis. Cai-Nicolas Ziegler et. al \cite{ziegler2005improving} propose a method to diversify personalized recommendation lists in order to reflect the user's complete spectrum of interests. They achieve this by introducing an intra-list similarity metric to assess the topical diversity of  recommendation lists and then, reduce the intra-list similarity thereby diversifying the topics. This has shown to improve user satisfaction. \\

Specific to our problem of recommending conferences to authors, not much has been done in the dimensionality reduction space. The work done by H. Luong et al. \cite{luong2012publication} makes use of the social network of the authors. For every author of the paper in need of a conference recommendation, based on his/her social network, the various conferences are given weights. These are then combined and the conference with the highest weight is suggested for the paper. Zaihan Yang and Brian D. Davidson \cite{yang2012venue} provide a collaborative filtering-based recommender system that can provide venue recommendations to researchers. Here, papers are represented by both content (using topics, requiring LDA) and stylometric features. Eric Medvet et. al \cite{medvetpublication}, in their work, address the same problem but make use of only the title of the paper and abstract. They propose different approaches where they match the topics of a scientific paper with those of the possible publication venues for that paper.\\

We make use of Correspondence Analysis in our work, building a model out of the content of the abstracts thereby leading to a more meaningful conference recommender system for papers. We employ dimensionality reduction techniques to further reduce the noise and result in better recommendations. In this way, our approaches are different from those that are previously attempted: dimensionality reduction techniques, to the best of our knowledge, have not been explored very extensively in the text-based recommendation space. Therein lies our novelty and contribution to the machine learning literature.\\

\subsection{Data Mining Techniques}
\label{subsec:data_mining_techniques}

Decision trees may be used in a model-based approach for a RS. One possibility is to use content features to build a decision tree that models all the variables involved in the user preferences. Bouza et al. \cite{bouza2008semtree} use this idea to construct a Decision Tree using semantic information available for the items. The tree is built after the user has rated only two items. The features for each of the items are used to build a model that explains the user ratings. They use the information gain of every feature as the splitting criteria. Another option to use Decision Trees in a RS is to use them as a tool for item ranking. The use of Decision Trees for ranking has been studied in several settings and their use in a RS for this purpose is fairly straightforward \cite{cheng2009decision}. \\

A rule-based system can be used to improve the performance of a RS by injecting partial domain knowledge or business rules. Gutta et al. \cite{gutta2000tv} implemented a rule-based RS for TV content. In order to do, so they first derived a $C4.5$ Decision Tree that is then decomposed into rules for classifying the programs. \\

Bayesian classifiers are particularly popular for model-based RS. They are often used to derive a model for content-based RS. However, they have also been used in a CF setting. Miyahara and Pazzani \cite{miyahara2000collaborative} implement a RS based on a Naive Bayes classifier. In order to do so, they define two classes: like and don’t like. Experiments show that this model performs better than a correlation-based CF. \\

Support Vector Machines have recently gained popularity for their performance and efficiency in many settings. SVM’s have also shown promising recent results in RS. Kang and Yoo \cite{yoo2007svm}, for instance, report on an experimental study that aims at selecting the best preprocessing technique for predicting missing values for an SVM-based RS. \\

Xue et al. \cite{xue2005scalable} present a typical use of clustering in the context of a RS by employing the $k$-means algorithm as a pre-processing step to help in neighborhood formation. \\

Cho et al. \cite{cho2002personalized} combine Decision Trees and Association Rule Mining in a web shop RS. In their system, association rules are derived in order to link related items. The recommendation is then computed by intersecting association rules with user preferences. They look for association rules in different transaction sets such as purchases, basket placement, and click-through. They also use a heuristic for weighting rules coming from each of the transaction sets. Purchase association rules, for instance, are weighted higher than click-through association rules. \\

Recently, several approaches involving natural language processing \cite{iyer2019event,iyer2019unsupervised,iyer2019heterogeneous,iyer2017detecting,iyer2019machine,iyer2017recomob,iyer2019simultaneous}, machine learning \cite{li2016joint,iyer2016content,honke2018photorealistic}, deep learning \cite{iyer2018transparency,li2018object} and numerical optimizations \cite{radhakrishnan2016multiple,iyer2012optimal,qian2014parallel,gupta2016analysis,radhakrishnan2018new} have also been used in the visual and language domains.

\section{Problem Statement}
\label{sec:problem_statement}
Selection of publishing venue for a research work is an arduous task. With huge number of venues to choose from researchers may find it difficult to filter the appropriate conference for their paper. Hence, we try to automate the process of filtering and ordering the conferences. Let $\Phi = {p_1, p_2, \ldots, p_m}$ be the set of $m$ papers to be published, $\Gamma = {c_1, c_2, \ldots, c_n}$ be the set of $n$ conferences, and $u$ be the utility function such that, $u:\Phi \times \Gamma \rightarrow R$ where $R$ is a total ordered set. Then, we need to find the conference, $c_k \in \Gamma$ that maximizes the utility, $u$ for paper $p_z \in \Phi$.

\section{Preliminary Concepts}
\label{sec:preliminary_concepts}
\subsection{Correspondence Analysis}
\label{subsec:correspondence_analysis}
This section describes Correspondence Analysis in detail. The following section describes the theoretical aspect and the section after that discusses the computational details.
\subsubsection{Theory}
\label{subsubsec:theory}
Correspondence analysis (CA) is a multivariate statistical technique applied to categorical data usually in the form of a contingency table, rather than continuous data as in the case of PCA, and represents graphically the row and column categories thereby allowing for a comparison of their \textit{correspondences} or associations at a category level. CA tries to identify components in the reduced dimension to maximize the relations among the variables while PCA tries to get components that maximize the variability.\\

\paragraph{Introduction}
\label{subsubsubsec:introduction}

\begin{table}
\vspace{2ex}
\begin{tabular}{l | c | c | c | c | c}
\toprule
 & \multicolumn{5}{c}{\textbf{Hair Colour}} \\\cmidrule{2-6}
\textbf{Eye Colour}	& Fair & Red & Medium & Dark & Black \\
\midrule
Blue & 326 & 38 & 241 & 110 & 3 \\\specialrule{\cmidrulewidth}{1pt}{1pt}
Light & 688 & 116 & 584 & 188 & 4 \\\specialrule{\cmidrulewidth}{1pt}{1pt}
Medium & 343 & 84 & 909 & 412 & 26 \\\specialrule{\cmidrulewidth}{1pt}{1pt}
Dark & 98 & 48 & 403 & 681 & 85 \\
\bottomrule
\end{tabular}
\centering
\caption{Two-way contingency table classifying 5387 children in Caithness, Scotland, according to hair colour and eye colour \cite{beh2004simple}\label{contingency}}
\end{table}

A contingency table is a type of table in a matrix format that displays the frequency distribution of the variables. An example of one is shown in Table ~\ref{contingency}. A contingency table is usually associated with a \textit{grand total}, the total number of entities represented in the table, and \textit{marginal totals}, which are the row sums and column sums.\\

Correspondence analysis basically tries to find out any possible relation between the categorical variables. Contingency tables with more number of variables are possible, but they become difficult to visualize. So, analysis of contingency tables with only two variables are described here (the size of the grid can be anything).\\

The most basic concept in CA is that of a \textit{profile}, which is a set of frequencies divided by their total. For a given contingency table, we can have row and column profiles. The objective of CA is to be able to visualize these profiles and the relationships among them (for example the relation between hair colour and eye colour in Table ~\ref{contingency}), by projecting them onto a subspace of low dimensionality which best fits the profiles and the loss of information is minimized. Since the objective of finding low-dimensional best-fitting subspaces coincides with the objective of finding low-rank matrix approximations by least-squares, the SVD forms the backbone of CA. On a side note, CA is symmetrical in nature i.e. column analysis and row analysis yield the same results.\\

It so happens that row profiles of $J$ dimensions (meaning that the contingency table in consideration has $J$ columns), on being plotted in $J$ dimensions, lie on a $J-1$ dimensional space. This means that for a table with $3$ columns, the rows (after being divided by the row total to get profiles) lie on a $2$-D space, which is a plane. A similar situation arises with column profiles too: considering there are $I$ rows in the table, the column profiles lie in a $I-1$ dimensional subspace. Since the analysis is symmetrical, it can be observed that both the columns and rows have to lie in a min ($I-1$, $J-1$) dimensional subspace. \\

For a contingency table, the column categories can be thought to be a ``pure" row profile, i.e. it's distribution in the other column categories is $0$. So, the row profile vector of the second column category can be thought to be $[0, 1, 0, \ldots, 0]$. These points will form the vertices of the min${I-1, J-1}$ dimensional subspace that the row profiles lie in. Upon reducing the dimensions of both the row profiles and vertices using SVD, so that minimum information is lost by finding the \textit{principal axes}, it can often be brought down to a 2-D plot where the relationships are easily visualized. \\

The coordinates of the row profiles in the reduced subspace are called \textit{row principal coordinates} and the those of the vertices are called \textit{column standard coordinates}. A similar explanation can be given for \textit{column principal coordinates} and \textit{row standard coordinates}. Algorithm to compute these are given in section \ref{subsubsec:implementationCA}.\\

Having computed the principal and standard coordinates, the original information can be obtained back with some loss. This is called the \textit{reconstitution formula}:
\begin{equation}
	p_{ij} = r_i c_j \left(1 + \sum\limits_{k=1}^K \sqrt{\lambda_k} \phi_{ik} \gamma_{jk}\right)
\end{equation}

where
\begin {itemize}
	\item $p_{ij}$ are the relative proportions $n_{ij}/n$, $n$ being the grand total $\sum\nolimits_{i}$ $\sum\nolimits_{j}$ $n_{ij}$
	\item $r_i$ and $c_j$ are the row and column masses respectively
	\item $\lambda_k$ is the $k$-$th$ principal inertia
	\item $\phi_{ik}$ and $\gamma_{jk}$ are the row and column standard coordinates respectively
\end{itemize}
In the summation, there are as many terms $K$ as there are dimensions in the data matrix, which has been shown to be equal to one less than the number of rows or columns, whichever is smaller. Taking lesser dimensions than this will lead to some loss of information. We have to choose $K$ appropriately to minimize the loss.\\

In order to visualize the relationships between the categories, the coordinates of the rows and columns are plotted in a $2$-D map called \textit{biplot}. There are difference kinds of plots: \textit{asymmetrical} where the principal coordinates of rows/columns and the standard coordinates of the other are plotted. This is the most common representation to measure distances between the points to measure relationships etc. Another kind of plot is called \textit{symmetrical}, where both the rows and columns being depicted use the same coordinates: principal or standard.\\

\subsubsection{Theoretical Development}
\label{subsubsec:theoretical_dev}
\paragraph{Pearson's Chi-square Test for Independence}
\label{subsubsubsec:pearson_independence}

The Chi-square test is intended to test how likely it is that an observed distribution is due to chance. It is also called a ``goodness of fit" statistic, because it measures how well the observed distribution of data fits with the distribution that is expected if the variables are independent. A Chi-square test is designed to analyze categorical data, i.e. the data has been counted and divided into categories, and will not work with parametric or continuous data.\\

The Chi-square test basically tests the null hypothesis that the variables are independent. The test compares the observed data to a model that distributes the data according to the expectation that the variables are independent. Wherever the observed data doesn't fit the model, the likelihood that the variables are dependent becomes stronger, thus proving the null hypothesis incorrect. So, a Chi-square test would allow us to test how likely it is that the attendance state and outcome state are completely independent. The Chi-square test is only meant to test the probability of independence of a distribution of data. It does not give any details about the relationship between them. However, once the probability that the two variables are related is determined (using the Chi-square test), other methods can be used to explore their interaction in more detail.\\

To test the null hypothesis, we need to construct a model which estimates how the data should be distributed if our hypothesis of independence is correct. We build the required model by making use of the marginal and grand totals. The estimated value for each cell $(i, j)$, $E_{i, j}$ is given by
\begin{equation}
E_{i, j} = \frac{Row_i \; sum \times Column_j \; sum}{Grid \; total}
\end{equation}

This way, we get a table similar to the observed table, except that in this case, the variables are assumed to be independent. This table is used to test the null hypothesis by computing the \textit{Chi-square statistic}, $\chi^{2}$ as follows:
\begin{equation}
\chi^{2} = \sum\limits_{i = 1}^I \sum\limits_{j = 1}^J \frac{(O_{i, j} - E_{i, j})^2}{E_{i, j}}
\end{equation}

where $I$ represents the number of rows in the table and $J$ represents the number of columns in the table. $O_{i, j}$ stands for the $(i, j)^{th}$ entry of the observed table and $E_{i, j}$ stands for the expected/estimated value of the $(i, j)^{th}$ entry of the model table assuming independence.\\

Now, having calculated the Chi-square statistic, the numbers don't give much meaning unless we determine the \textit{p-value}. The \textit{p-value} is the probability of obtaining a test statistic result at least as extreme as the one that was actually observed, assuming that the null hypothesis is true.\\

With the Chi-square value and the \textit{degrees of freedom}, the p-value can be calculated. The degrees of freedom gives us the number of entries in the grid that are \textit{actually} independent. For a Chi-square grid, the degrees of freedom can be said to be the number of cells that need to be filled, given the totals in the margins, before the rest of the grid can be filled using a formula that depends on the marginal totals and the values in the cells filled earlier. Thus, for a Chi-square grid, the degrees of freedom are $(I - 1) \times (J - 1)$, where $I$ and $J$ represent the number of rows and columns in the table respectively.\\

\paragraph{Notation}
\label{subsubsubsec:notation}
Consider an $I \times J$ two-way contingency table $N$, where the $(i, j)^{th}$ cell entry is given by $n_{ij}$ for $i = 1, 2, \ldots, I$ and $j = 1, 2, \ldots, J$. Let the grand total of $N$ be $n$ and the correspondence matrix or matrix of relative frequencies be $P$ so that the $(i, j)^{th}$ cell entry is $p_{ij} = \sfrac{n_{ij}}{n}$ and $\sum\limits_{i = 1}^I \sum\limits_{j = 1}^J p_{ij} = 1$. Define the $i^{th}$ row marginal proportion by $p_i = \sum\limits_{j = 1}^J p_{ij}$ and define the $j^{th}$ column marginal proportion by $p_j = \sum\limits_{i = 1}^I p_{ij}$.

\paragraph{Pearson's Ratio}
\label{subsubsubsec:pearson}
The aim of correspondence analysis, like many multivariate data analytic techniques, is to determine scores which describe how similar or different responses to two or more variables are.\\

If we consider a model of complete independence between rows and columns of the table, then
\begin{equation}
p_{ij} = p_i \times p_j
\end{equation}

But this complete independence is almost never satisfied. So, we introduce a constant $\alpha_{ij}$ such that the new relation becomes
\begin{equation}
p_{ij} = \alpha_{ij} p_i p_j
\end{equation}

As can be seen, if $\alpha_{ij} = 1$ for $i = 1, 2, \ldots, I$ and $j = 1, 2, \ldots, J$, then complete independence in the model is observed. Since, complete independence is seldom observed, the elements for which $\alpha_{ij} \neq 1$ by calculating
\begin{equation}
\alpha_{ij} = \frac{p_{ij}}{p_i \times p_j}
\end{equation}

Using the Pearson's ratio, the Pearson Chi-square statistic can be written as
\begin{equation}
\label{chisquared}
\chi^{2} = n \sum\limits_{i = 1}^I \sum\limits_{j = 1}^J p_i p_j (\alpha_{ij} - 1)^2
\end{equation}

This has a chi-squared distribution with $(I - 1)(J - 1)$ degrees of freedom.\\

A property of the Pearson chi-squared statistic is that as $n$ increases, so too does the statistic. This can hinder tests of association in the contingency tables. To overcome this problem, simple correspondence analysis considers $\sfrac{\chi^2}{n}$, which is referred to as the \textit{total inertia} of the contingency table, to describe the level of association, or dependence, between two categorical variables.
By decomposing the total inertia, important sources of information that help describe this association can be identified. Most commonly, SVD is used to decompose the Pearson's ratio.

\paragraph{Using Singular Value Decomposition}
\label{subsubsubsec:svd}
Classically, simple correspondence analysis is conducted by performing a singular value decomposition (SVD) on the Pearson ratio. The method of SVD, also referred to as the “Eckart Young” decomposition, is the most common tool used to decompose the Pearson ratio. For the application of analysis of contingency tables, the Pearson ratio may be decomposed into components by
\begin{equation}
\label{svd_decomposition}
\alpha_{ij} = \sum\limits_{m = 0}^{M^{*}} a_{im} \lambda_{m} b_{jm}
\end{equation}

where $M^{*} = min{I, J} - 1$ is the maximum number of dimensions required to graphically depict the association between the row and column responses. For example, for Table \ref{contingency} only $min{4, 6} - 1 = 3$ dimensions are required to graphically depict all of the association between the hair and eye colour of the children classified in Caithness. However, for a simple interpretation of this association, generally only the first two dimensions are used to construct such a graphical summary.\\

The vector \textbf{$a_m$}$ = (a_{1m}, a_{2m}, \ldots, a_{Im})$ is the $m^{th}$ row singular vector and is associated with the $I$ row categories. Similarly, \textbf{$b_m$}$ = (b_{1m}, b_{2m}, \ldots, b_{Jm})$ is the $m^{th}$ column singular vector and is associated with the $J$ column categories. The elements of the vector \textbf{$\lambda$}$ = (\lambda_0, \lambda_1, \ldots, \lambda_M^{*})$ are real and positive and are the first $M^{*}$ singular values and are arranged in descending order so that
\begin{equation}
\lambda_0 = 1 \geq \lambda_1 \geq \ldots \geq \lambda_M^{*} \geq 0
\end{equation}

These singular values can be also be calculated by
\begin{equation}
\lambda_m = \sum\limits_{i = 1}^I \sum\limits_{j = 1}^J a_{im} b_{jm} p_{ij}
\end{equation}

while the singular vectors have the property
\begin{equation}
\label{ortho}
\sum\limits_{i = 1}^I p_i a_{im} a_{im'} = \begin{cases}
												1 & m = m' \\
												0 & m \neq m'
											\end{cases} \quad
\sum\limits_{j = 1}^J p_j b_{jm} b_{jm'} = \begin{cases}
												1 & m = m' \\
												0 & m \neq m'
											\end{cases}
\end{equation}

We use the fact that $\lambda_0 = 1$, $a_{i0} = 1$ and $b_{j0} = 1$ to rewrite equation (\ref{svd_decomposition}) as
\begin{equation}
\label{newalpha}
\alpha_{ij} = 1 + \sum\limits_{m = 1}^{M^{*}} a_{im} \lambda_{m} b_{jm}
\end{equation}

Using equation (\ref{newalpha}) in equation (\ref{chisquared}), we get
\begin{equation}
\begin{split}
\chi^{2} & = n \sum\limits_{i = 1}^I \sum\limits_{j = 1}^J p_i p_j (\alpha_{ij} - 1)^2 \\
	& = n \sum\limits_{i = 1}^I \sum\limits_{j = 1}^J p_i p_j (\sum\limits_{m = 1}^{M^{*}} a_{im} \lambda_{m} b_{jm})^2 \\
	& = n \sum\limits_{m = 1}^{M^{*}} \lambda_m^{2} (\sum\limits_{i = 1}^I p_i a_{im}^2)^2 (\sum\limits_{j = 1}^J p_j b_{jm}^2)^2
\end{split}
\end{equation}

By using the orthogonality properties of $a_{im}$ and $b_{jm}$ from equation (\ref{ortho}), the total inertia can be written in terms of singular values such that
\begin{equation}
\frac{\chi^2}{n} = \sum\limits_{m = 0}^{M^{*}} \lambda_m^2
\end{equation}

For Table \ref{contingency}, we obtain $\lambda_1^2 = 0.1992$, $\lambda_2^2 = 0.0301$, $\lambda_3^2 = 0.0009$ so that $\sfrac{\chi^2}{n} = 0.2302$. So, the first axis explains $\sfrac{0.1992}{0.2302} = 0.8656$ of the total variation that exists in the table, while the second axis explains $\sfrac{0.0301}{0.2302} = 0.1307$ of this variation. Thus, considering just these two axes accounts for $99.63\%$ of the total variation in Table \ref{contingency}. So, we can safely ignore the $3^{rd}$ component, i.e. corresponding to $\lambda_3$ without much loss. This way, we are reducing the dimensions.\\

\subsubsection{Implementation}
\label{subsubsec:implementationCA}
CA is based on fairly straightforward, classical results in matrix theory. The central result is the singular value decomposition (SVD), which is the basis of many multivariate methods such as principal component analysis, canonical correlation analysis, all forms of linear biplots, discriminant analysis and metric multidimensional scaling. Here, Matrix–vector notation is used because it is more compact \cite{greenacre2007correspondence}.\\

\noindent Let $\bm{N}$ denote the $I \times J$ data matrix with positive row and column sums. For notational simplicity, the matrix is first converted to the \textit{correspondence matrix} $\bm{P}$ by dividing $\bm{N}$ by it's grand total $n = \sum\nolimits_{i} \sum\nolimits_{j} n_{ij}$.\\

\noindent \textit{Correspondence Matrix: }
\begin{equation}
	\bm{P} = \frac{1}{n} \bm{N}
\end{equation}

\noindent The following notation is used:\\
\noindent \textit{Row and Column masses:}



\begin{equation}
	\begin{aligned}[c]
		r_i &= \sum\limits_{j=1}^J p_{ij}\\
		\bm{r} &= \bm{P}\bm{1}
	\end{aligned}
	\qquad\qquad
	\begin{aligned}[c]
		c_j &= \sum\limits_{i=1}^I p_{ij}\\
		\bm{c} &= \bm{P^{T}}\bm{1}
	\end{aligned}
\end{equation}

%


\noindent \textit{Diagonal matrices of row and column masses:}
\begin{equation}
  \bm{D}_r = diag(\bm{r}) \qquad\text{and}\qquad \bm{D}_c = diag(\bm{c})
\end{equation}

All subsequent results are given in terms of these relative quantities $\bm{P} = {p_{ij}}$, $\bm{r} = {r_i}$ and $\bm{c} = {c_j}$, whose elements add up to $1$ in each case.

\paragraph{Basic Computational Algorithm}
\label{subsubsubsec:basicCAcomp}
The computational algorithm to obtain coordinates of the row and column profiles with respect to principal axes, using the singular value decomposition (SVD), is as follows:\\
\begin{itemize}
	\item \textit{Step 1: Calculate the matrix} \textbf{S} \textit{of standardized residuals}
		\begin{equation}
			\bm{S} = \bm{D^{-\frac{1}{2}}}_r (\textbf{P} - \bm{rc^T}) \bm{D^{-\frac{1}{2}}}_c
		\end{equation}
	\item \textit{Step 2: Calculate the SVD of} \textbf{S}
		\begin{equation}
  		\bm{S} = \bm{U}\bm{D}_{\alpha}\bm{V^{T}} \;\;\text{where}\;\; \bm{U^{T}U} = \bm{V^{T}V} = \bm{I}
		\end{equation}
		where $\bm{D}_{\alpha}$ is the diagonal matrix of (positive) singular values in descending order: $\alpha_{1}$ $\geq$ $\alpha_{2}$ $\geq$ $\ldots$
	\item \textit{Step 3: Standard coordinates} \textbf{$\Phi$} \textit{of rows}
		\begin{equation}
			\bm{\Phi} = \bm{D^{-\frac{1}{2}}}_r \bm{U}
		\end{equation}
	\item \textit{Step 4: Standard coordinates} \textbf{$\Gamma$} \textit{of columns}
		\begin{equation}
			\bm{\Gamma} = \bm{D^{-\frac{1}{2}}}_c \bm{V}
		\end{equation}
	\item \textit{Step 5: Principal coordinates} \textbf{F} \textit{of rows}:
		\begin{equation}
			\bm{F} = \bm{\Phi} \bm{D}_{\alpha}
		\end{equation}
	\item \textit{Step 6: Principal coordinates} \textbf{G} \textit{of columns}:
		\begin{equation}
			\bm{G} = \bm{\Gamma} \bm{D}_{\alpha}
		\end{equation}
	\item \textit{Step 7: Principal inertias} $\lambda_{k}$:
		\begin{equation}
			\lambda_{k} = \alpha^{2}_k, \;\; k = 1, 2, \ldots, K \;\; \text{where} \;\; K = min \left\lbrace I-1,\; J-1\right\rbrace
		\end{equation}
\end{itemize}

\paragraph{Transition equations between rows and columns}
\label{subsubsubsec:transition_equations}
The left and right singular vectors are related linearly, for example by multiplying the SVD on the right by $\bm{V}: \bm{SV} = \bm{U}\bm{D}_{\alpha}$. Expressing such relations in terms of the principal and standard coordinates gives the following variations of the same theme, called transition equations:
\begin{itemize}
	\item \textit{Principal as a function of standard (barycentric relationships)}
		\begin{equation}
			\bm{F} = \bm{D}^{-1}_r \bm{P} \bm{\Gamma} \qquad \bm{G} = \bm{D}^{-1}_c \bm{P^T} \bm{\Phi}
		\label{suppcol}
		\end{equation}
	\item \textit{Principal as a function of principal}
		\begin{equation}
			\bm{F} = \bm{D}^{-1}_r \bm{P} \bm{G} \bm{D}^{-1/2}_{\lambda} \qquad \bm{G} = \bm{D}^{-1}_c \bm{P^T} \bm{F} \bm{D}^{-1/2}_{\lambda}
		\end{equation}
\end{itemize}

\paragraph{Supplementary Points}
\label{subsubsubsec:supplementary_points}
Supplementary rows/columns are those entries that are added to the original table. In many cases. we may require their principal/standard coordinates. The transition equations can be used to situate the supplementary points on the map. This way, we can compute the coordinates for the supplementary points using the already computed coordinates for the original table.

For example, given a supplementary column point with values in $\bm{h} \; (I \times 1)$, divide by its total $\bm{1^T h}$ to obtain the column profile $\tilde{\bm{h}}$, and then use the profile transposed as a row vector in the second equation of (\ref{suppcol}), for example, to calculate the coordinates \textbf{g} of the supplementary column
\begin{equation}
	\bm{g} = \tilde{\bm{h}}^{\bm{T}} \bm{\Phi}
\end{equation}

In the proposed methods, we required the principal coordinates of the supplementary rows (which are the rows of the test matrix). The steps taken to obtain those, from the trained model, are:
\begin{itemize}
	\item \textit{Step 1:} $\bm{N}_{test}$ is our test matrix
	\item \textit{Step 2:} We obtain the correspondence matrix $\bm{P}_{test}$ of $\bm{N}_{test}$, by normalizing the entries of the matrix with its grand total $n$
		\begin{equation}
			\bm{P}_{test} = \frac{1}{n} \bm{N}_{test}
		\end{equation}
	\item \textit{Step 3:} We obtain the row masses $\bm{r}_{test}$ for $\bm{P}_{test}$
		\begin{equation}
			\bm{r}_{test} = \bm{P}_{test}\bm{1}
		\end{equation}
	\item \textit{Step 4:} The diagonal matrix corresponding to $\bm{r}_{test}$, $\bm{D}_{r_{test}}$, is obtained
		\begin{equation}
			\bm{D}_{r_{test}} = diag(\bm{r}_{test})
		\end{equation}
	\item \textit{Step 5:} From the trained model, we have the column standard coordinates for the training matrix $\bm{N}_{train}$. Let us call that column standard coordinate matrix as $\bm{\Gamma}_{train}$. Then using the transition equations given in equation (\ref{suppcol}), we obtain the principal coordinates for the test matrix (supplementary rows), $\bm{F}_{test}$, as
		\begin{equation}
			\bm{F}_{test} = \bm{D}^{-1}_{r_{test}} \bm{P}_{test} \bm{\Gamma}_{train}
		\end{equation}
\end{itemize}

Using the procedures mentioned in this section, CA has been implemented and used for the experiments conducted. For more details about the procedures and implementations, one can refer to \cite{greenacre2007correspondence}.

\subsection{Topic Modeling}
\label{subsec:topic_modeling}
In machine learning and natural language processing, a topic model is a type of statistical model for discovering the abstract ``topics'' that occur in a collection of documents. Intuitively, given that a document is about a particular topic, one would expect particular words to appear in the document more or less frequently: ``dog'' and ``bone'' will appear more often in documents about dogs, ``cat'' and ``meow'' will appear in documents about cats, and ``the'' and ``is'' will appear equally in both. A document typically concerns multiple topics in different proportions; thus, in a document that is $10\%$ about cats and $90\%$ about dogs, there would probably be about $9$ times more dog words than cat words. A topic model captures this intuition in a mathematical framework, which allows examining a set of documents and discovering, based on the statistics of the words in each, what the topics might be and what each document's balance of topics is.

\subsubsection{Preliminaries}
\label{subsubsec:prelims}
The model being looked at is the \textit{Latent Dirichlet Allocation (LDA)}, which is a \textit{bag-of-words} model.

\paragraph{Bag-of-words Model}
\label{subsubsubsec:bow}
\begin{figure}
\centering
\includegraphics[scale=0.5]{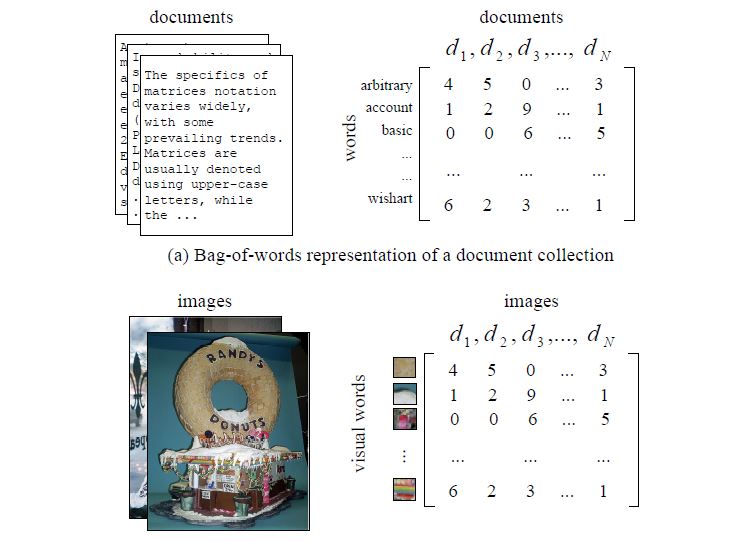}
\caption{Understanding \textit{Bag-of-words} representation \cite{Xiao2011}}
\label{bow}
\end{figure}

This model is a simplifying assumption used in natural language processing and information retrieval wherein, a text (such as a sentence or a document) is represented as an unordered collection of words, disregarding grammar and even word order. From a data modeling point of view, the bag-of-words model can be represented by a co-occurrence matrix of documents and words as illustrated in Figure ~\ref{bow}. Just as a text consists of words, a multimedia document can be thought to consist of \textit{sensory words}, thus allowing them a bag-of-words representation too. This model is widely used in document classification and modeling. When a Naive Bayes classifier is applied to text, for example, the conditional independence assumption leads to the bag-of-words model. Other methods of document modeling that use this model include the Latent Dirichlet Allocation.

\subsubsection{Latent Dirichlet Allocation (LDA)}
\label{subsubsec:lda}
\begin{figure}
\centering
\includegraphics[scale=0.8]{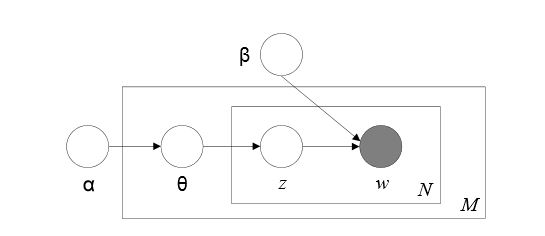}
\caption{Graphical model representation of LDA. The boxes are \textit{plates} representing replicates. The outer plate represents documents, while the inner plate represents the repeated choice of topics and words within a document \cite{blei2003latent}.}
\label{lda}
\end{figure}

LDA is a generative probabilistic model that generates a document using a mixture of topics \cite{blei2003latent}. It assumes a generative probabilistic model in which documents are represented as random mixtures over latent topics, where each topic is characterized by a probability distribution over words. An illustration of the assumption in LDA model is depicted in Figure ~\ref{lda}.\\

LDA\cite{blei2003latent} assumes the following generative process for each document \textbf{w} in a corpus $D$:
\begin{enumerate}
\item Choose $N$ $\sim$ Poisson($\xi$)
\item Choose $\Theta$ $\sim$ Dir($\alpha$)
\item For each of the $N$ words $w_n$:
	\begin{itemize}
		\item Choose a topic $z_n$ $\sim$ Multinomial($\Theta$)
		\item Choose a word $w_n$ from $p(w_n|z_n, \beta)$, a multinomial probability conditioned on the topic $z_n$
	\end{itemize}
\end{enumerate}

\begin{figure}
\centering
\includegraphics[scale=0.5]{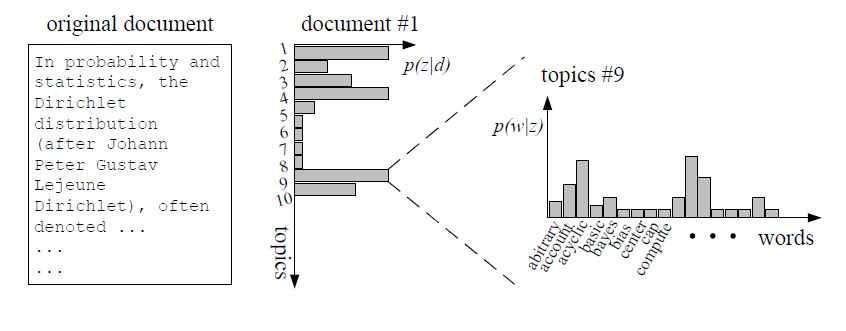}
\caption{Visualization of LDA \cite{Xiao2011}.}
\label{lda-vis}
\end{figure}

A visualization of LDA is given in Figure ~\ref{lda-vis}.

\paragraph{Understanding LDA with an example}
\label{subsubsubsec:understanding_lda}
Suppose we have the following sentences:

\begin{itemize}
	\item I ate a banana and spinach smoothie for breakfast
	\item I like to eat broccoli and bananas.
	\item Chinchillas and kittens are cute.
	\item My sister adopted a kitten yesterday.
	\item Look at this cute hamster munching on a piece of broccoli.
\end{itemize}

Latent Dirichlet allocation is a way of automatically discovering topics that these sentences contain. For example, given these sentences and asked for 2 topics, LDA might produce something like

\begin{itemize}
	\item \textbf{Sentences 1 and 2}: 100
  \item \textbf{Sentences 3 and 4}: 100
  \item \textbf{Sentence 5}: 60
  \item \textbf{Topic A}: 30
  \item \textbf{Topic B}: 20
\end{itemize}

The question, of course, is: how does LDA perform this discovery?\\

In more detail, LDA represents documents as mixtures of topics that spit out words with certain probabilities. It assumes that documents are produced in the following fashion: when writing each document, we

\begin{itemize}
 	\item Decide on the number of words N the document will have (say, according to a Poisson distribution).
	\item Choose a topic mixture for the document (according to a Dirichlet distribution over a fixed set of K topics). For example, assuming that we have the two food and cute animal topics above, you might choose the document to consist of 1/3 food and 2/3 cute animals.
	\item Generate each word in the document by:
		\begin{itemize}
    	\item First picking a topic (according to the multinomial distribution that you sampled above; for example, you might pick the food topic with 1/3 probability and the cute animals topic with 2/3 probability).
    	\item Then using the topic to generate the word itself (according to the topic's multinomial distribution). For instance, the food topic might output the word ``broccoli" with 30\% probability, ``bananas" with 15\% probability, and so on.
    \end{itemize}

Assuming this generative model for a collection of documents, LDA then tries to backtrack from the documents to find a set of topics that are likely to have generated the collection.
\end{itemize}

As an example, according to the above process, when generating some particular document $D$, we might
\begin{itemize}
	\item Decide that D will be 1/2 about food and 1/2 about cute animals.
  \item Pick 5 to be the number of words in D.
  \item Pick the first word to come from the food topic, which then gives you the word ``broccoli".
  \item Pick the second word to come from the cute animals topic, which gives you ``panda".
  \item Pick the third word to come from the cute animals topic, giving you ``adorable".
  \item Pick the fourth word to come from the food topic, giving you ``cherries".
  \item Pick the fifth word to come from the food topic, giving you ``eating".
\end{itemize}

So the document generated under the LDA model will be ``broccoli panda adorable cherries eating" (note that LDA is a bag-of-words model). Now, how to infer the parameters of the LDA model, given a set of documents, is described in the next section.

\paragraph{Learning}
\label{subsubsubsec:learning}
So now suppose we have a set of documents. We have chosen some fixed number of topics, $K$, to discover, and want to use LDA to learn the topic representation of each document and the words associated to each topic. In short, we want to perform an inference on this generative model. Several techniques like EM algorithm, Gibbs Sampling etc. can be used for this purpose. For our experiments, we have used \textit{collapsed Gibbs sampling}:

\begin{enumerate}
\item Go through each document, and randomly assign each word in the document to one of the $K$ topics.
\item Notice that this random assignment already gives us both topic representations of all the documents and word distributions of all the topics (albeit not very good ones).
\item So to improve on them:
	\begin{itemize}
		\item For each document $d$
		\item Go through each word $w$ in $d$
		\item And for each topic $t$, compute two things: 1) $p(\text{topic} \; t \, | \, \text{document} \; d)$ = the proportion of words in document $d$ that are currently assigned to topic $t$, and 2) $p(\text{word} \; w \, | \, \text{topic} \; t)$ = the proportion of assignments to topic $t$ over all documents that come from this word $w$. Reassign $w$ a new topic, where we choose topic $t$ with probability $p(\text{topic} \; t \, | \, \text{document} \; d) \times p(\text{word} \; w \,| \, \text{topic} \; t)$ (according to our generative model, this is essentially the probability that topic $t$ generated word $w$, so it makes sense that we resample the current word's topic with this probability).
		\item In other words, in this step, we're assuming that all topic assignments except for the current word in question are correct, and then updating the assignment of the current word using our model of how documents are generated.
	\end{itemize}
\item After repeating the previous step a large number of times, we will eventually reach a roughly steady state where our assignments are pretty good. So, we can use these assignments to estimate the topic mixtures of each document (by counting the proportion of words assigned to each topic within that document) and the words associated to each topic (by counting the proportion of words assigned to each topic overall).
\end{enumerate}

\subsection{Term Frequency-Inverse Document Frequency}
\label{subsec:tf-idf}
One of the best-known measures for specifying keyword weights in Information Retrieval is the term frequency/inverse document frequency (TF-IDF) measure that is defined as follows: Assume that $N$ is the total number of documents that can be recommended to users and that keyword $k_i$ appears in $n_i$ of them. Moreover, assume that $f_{i,j}$ is the number of times keyword $k_i$ appears in document $d_j$. Then, $TF_{i,j}$, the term frequency (or normalized frequency) of keyword $k_i$ in document $d_j$, is defined as
\begin{equation}
TF_{i,j} = \frac{f_{i,j}}{max_z f_{z, j}}
\end{equation}

where the maximum is computed over the frequencies $f_{z,j}$ of all keywords $k_z$ that appear in the document $d_j$. However, keywords that appear in many documents are not useful in distinguishing between a relevant document and an irrelevant one. Therefore, the measure of inverse document frequency ($IDF_i$) is often used in combination with simple term frequency ($TF_{i,j}$). The inverse document frequency for keyword $k_i$ is usually defined as
\begin{equation}
IDF_i = \log \frac{N}{n_i}
\end{equation}

Then, the TF-IDF weight for keyword $k_i$ in document $d_j$ is defined as

\begin{equation}
w_{i,j} = TF_{i,j} \times IDF_i
\end{equation}

and the content of document $d_j$ is defined as
\begin{equation}
Content(d_j) = (w_{1j}, \ldots, w_{kj})
\end{equation}

which is a vector of weights.

We use this, LDA and CA in our experiments: tf-idf, LDA to represent content and CA to reduce the dimensions.

\section{Data Set and Tools Used}
\label{sec:data_sets_tools}
\subsection{Data Used}
\label{subsec:data_used}
Techniques based on the network analysis of authors and content analysis of the publications, have been explored for the purposes of recommendation. Each of the following subsections describe the data collected and techniques/tools applied on the data. For uniformity, we have used the publications in ACM conferences over the years 2008 to 2010. The selected conferences include

\begin{enumerate}
\item SIGBED - Special Interest Group on Embedded System
	\begin{itemize}
		\item CASES - Compilers, Architecture, and Synthesis for Embedded Systems
		\item CODES + ISSS - International Conference on Hardware/Software Codesign and Systems Synthesis
		\item EMSOFT - International Conference on Embedded Software
		\item SENSYS - Conference On Embedded Networked Sensor Systems
	\end{itemize}
\item SIGDA - Special Interest Group on Design Automation
	\begin{itemize}
		\item DAC - Design Automation Conference
		\item DATE - Design, Automation, and Test in Europe
		\item ICCAD - International Conference on Computer Aided Design
		\item SBCCI - Annual Symposium On Integrated Circuits And System Design
	\end{itemize}
\item SIGIR - Special Interest Group on Information Retrieval
	\begin{itemize}
		\item CIKM - International Conference on Information and Knowledge Management
		\item JCDL - ACM/IEEE Joint Conference on Digital Libraries
		\item SIGIR - Research and Development in Information Retrieval
		\item WWW - World Wide Web Conference Series
	\end{itemize}
\item SIGPLAN - Special Interest Group on Programming Languages
	\begin{itemize}
		\item GPCE - Generative Programming and Component Engineering
		\item ICFP - International Conference on Functional Programming
		\item OOPSLA - Conference on Object-Oriented Programming Systems, Languages, and Applications
		\item PLDI - Programming Language Design and Implementation
	\end{itemize}
\end{enumerate}

All together there are 16 conferences, which are from the 4 special interest groups. SIGBED is special interest group on embedded systems and accepts contributions related to embedded computer systems including software and hardware. SIGDA is special interest group on design automation. It accepts contributions on design and automation of complex systems on chip. SIGIR accepts contributions related to any aspect of Information Retrieval (IR) theory and foundation, techniques and applications. SIGPLAN is special interest group on programming languages and accepts contributions related to design, implementation and principles of programming languages.

\subsubsection{Co-Author Network}
\label{subsubsec:co-author}
We have downloaded the DBLP database, which contains the conference proceedings. This database contains the XML records of all the publications. Each record contains its publication information such as: author names, publication venue, title, year, and the DOI (Digital Object Identifier) of the publication. We extracted these attributes and generated a co-author network. Each node in the co-author network represents an author and each edge represents the co-authorship between the author nodes.

\subsubsection{Content-Analysis}
\label{subsubsec:content-analysis}
The ACM site provides abstracts for all the publications on its website. In order to perform content analysis, we crawled the ACM site and extracted the abstracts over the years 2008 to 2010 from the above mentioned conferences. We extracted a total of about 5447 abstracts published in these conferences and used them for content-based recommendations.

\subsection{Tools Used}
\label{subsec:tools_used}
\subsubsection{Neo4j Graph Database}
\label{subsubsec:neo4j}
For constructing the co-author network, Neo4j graph database has been used. It is an open-source project for graph databases. The python bindings were used to interact with the database.

\subsubsection{MySQL Database}
\label{subsubsec:mysql_data}
We relied on MYSQL to store the information on publications like year, DOI and venue.

\subsubsection{Programming Languages}
\label{subsubsec:programming_language}
Latent Dirichlet Allocation (LDA) was written in C++. All the other applied methods were written in Python and R.

\section{Technical Approach}
\label{sec:work_done}
Different approaches can be taken to solve the considered problem of attempting to recommend conferences to authors. Outline of ideas are provided and their pitfalls, if any, are mentioned. This recommender system unlike most commercial ones like recommending books, movies etc. involves people in some sense. Thus, there is an emotional connection involved. What this means is, if a conference suggested by our system gets a paper rejected, it is highly unlikely that he will use this system again. This is not that case with books or movie recommenders. So, there is no room for errors and less accuracies.\\

Some previous work on this has been done by H. Luong et al. \cite{luong2012publication} who have recommended conferences to authors using the social network i.e. the co-author network with the same dataset. Exploring the possibility of using CA has not been attempted before.\\

We have implemented a total of $6$ methods for this application and have done a comprehensive evaulation of the results. Three of the methods use Correspondence Analysis and three of them don't. The first method uses the Author-conference relation without taking into account the content of the paper. The next two methods use the content along with an application of CA to arrive at the results. The abstracts of the paper are used for content-analysis. This makes sense because the essence of the entire paper is contained in the abstract. \\

The last three methods are respectively: Content-based filtering, Collaborative filtering and Hybrid filtering. Content-based filtering and Hybrid filtering use the content of the paper but none of these methods employ CA. \\

Content is obtained in two ways: term frequency-inverse document frequency (tf-idf) and topics. LDA has been used for the latter. For each content-method, number of topics used: $100$, $200$, $400$, $600$, $800$ and $1000$. However, only results for $400$ topics are displayed in the evaluation, due to there being a very vast multitude of results and it would be too cumbersome to list all of them. Number of words used in tf-idf: $14082$. For computing the resultant conferences, three methods of similarity have been used: euclidean distance, cosine similarity and pearson correlation.\\

In all the methods, $2008-2009$ set of papers have been used for training and $2010$ papers have been used for testing. There are a total of $5447$ papers for the years $2008-2010$, $3572$ for $2008-2009$ and $1875$ for $2010$. There are a total of $16$ conferences.\\

The various similarity metrics used in the experiments are given below:
\begin{itemize}
\item Euclidean distance:

\begin{equation}
d(x, y) = \sqrt{\sum\limits_{k = 1}^n (x_k - y_k)^2}
\end{equation}

\noindent where $n$ is the number of attributes and $x_k$ and $y_k$ are the $k^{th}$ attributes of the data points $x$ and $y$, respectively.

\item Cosine Similarity:\\
In this similarity measure, items are considered as n-dimensional document vectors and their similarityis measured as the cosine of the angle that they form between them. Thus, if the cosine measure is close to $1$, i.e. the angle between the two vectors is close to $0$, the items are considered to be very similar.

\begin{equation}
\label{cosinesim}
\cos(x, y) = \frac{(x \cdot y)}{\norm{x} \norm{y}}
\end{equation}

\noindent where $\cdot$ indicates vector dot product and $\norm{x}$ is the norm of vector x. This similarity is also known as the $L_2$ Norm.\\


\item Pearson Correlation: \\
Correlation between items can also measure their similarity, linear relationship in this case. Although several correlation coefficients can be used, the most commonly used one is the \textit{Pearson Correlation}. Given the covariance of data points $x$ and $y$, $\Sigma$, and their standard deviation $\sigma$, we compute the Pearson correlation using:

\begin{equation}
\label{pearson}
Pearson(x, y) = \frac{\Sigma(x, y)}{\sigma_x \times \sigma_y}
\end{equation}
\end{itemize}

\subsection{Using Authors-Conferences Matrix}
\label{subsec:author_conf_matrix}
\subsubsection{Data Construction}
\label{subsubsec:data_construction}
\begin{figure}
\centering
\includegraphics[scale=0.6]{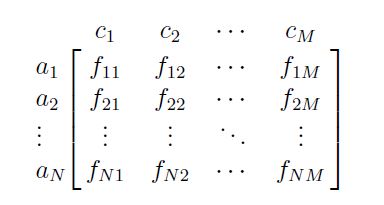}
\caption{The Author-Conference matrix}
\label{authorconf}
\end{figure}

From the data collected in the DBLP database, we construct the \textit{author-conference} matrix as shown in Figure \ref{authorconf}, where each row represents a single author. Here $f_{ij}$ represents the number of times author $a_i$ has published in conference $c_j$. We construct two such matrices: one training, say $\bm{M}_{train}$ and the other a test matrix, say $\bm{M}_{test}$. The training matrix $\bm{M}_{train}$ is constructed from $2008-2009$ papers (a total of $3572$) and the test matrix $\bm{M}_{test}$ is constructed from the $2010$ papers (a total of $1875$). There are a total of $16$ conferences. \\

\subsubsection{Applied Method}
\label{subsubsec:applied_method}
The algorithm followed is given in the following steps:

\begin{itemize}
	\item We compute the standardized residual matrix $\bm{S}_{train}$ from $\bm{M}_{train}$ as mentioned in section \ref{subsubsubsec:basicCAcomp}.
	\item We then obtain the coordinate matrices (both standard and principal for rows and columns), after decomposing $\bm{S}_{train}$ using SVD.
	\item Using the matrix $\bm{M}_{test}$ as a supplementary row matrix, we compute its principal coordinates using the standard column coordinates of $\bm{M}_{train}$.
	\item The rows of the supplementary test matrix $\bm{M}_{test}$ represent individual authors. So, to recommend a conference to a paper, which may be written by multiple authors: we take all the authors of that particular paper and compute the similarity (euclidean distance/cosine/pearson) with each of the $16$ conferences. For this purpose, we use the principal coordinates of the authors and the principal coordinates of the conferences.
	\item We sort the conferences, which maximize the sum of the similarity to all the authors of the paper in consideration, in decreasing order. Maximizing similarity means: minimizing euclidean distance/maximizing cosine similarity/maximizing pearson correlation.
	\item We then get a ranked list of recommendations for each paper.

\end{itemize}



This method has several drawbacks. For one, all the new authors (new to these conferences) are all recommended the same conference. Thus, this approach fails if the author has no publication history. Also, this does not capture the essence of the problem because we are recommending without even looking at the content of the paper in question. Thus, we need to look at the content of the paper as well in order to make better and more appealing recommendations.\\

Here, we have considered each row to be a single author. It can also be changed to comprise of multiple authors i.e. who have co-authored a paper. In this case, there will be more number of entries in the matrix and also it will be more sparse. Even in this case, the same limitations as above apply and in addition, the sparsity, in some sense, also reduces the ``meaningfulness" between the authors and conferences. Applying a dimensionality reduction technique like SVD or CA will bring it to a lower-dimensional subspace which will capture the essence of the relation better, rendering the matrix less sparse.

\subsection{Composition of Papers-Words/Topics and Words/Topics-Conferences Matrices}
\label{subsec:composition_topics}
\subsubsection{Data Construction}
\label{subsubsec:comp_data_construction}
\begin{figure}
\centering
\includegraphics[scale=0.6]{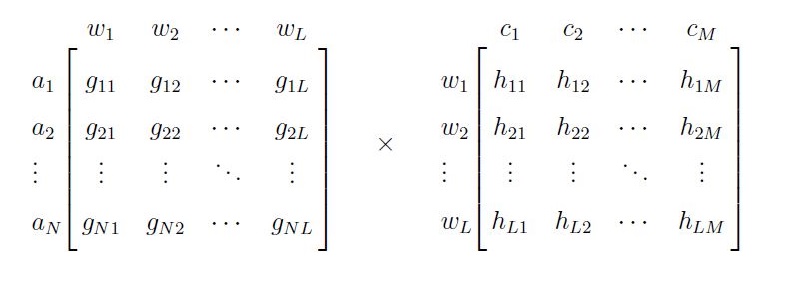}
\caption{The Paper-Word and Word-Conference Matrices}
\label{mult}
\end{figure}

A way to remedy the defect in the previous method is to look at the content of the papers, abstracts in particular as they capture the entire essence of the paper. From the data collected, we can construct an \textit{paper $\times$ words/topics} matrix and \textit{words/topics $\times$ conferences} matrix as shown in Figure \ref{mult}. We construct three matrices in total: two for training, and one for testing. We construct two training matrices, textit{paper $\times$ words/topics} and \textit{words $\times$ topics-conferences} from the $2008-2009$ papers, say $\bm{A}_{train}$ ($3572 \times 14082$) and $\bm{C}_{train}$ ($14082 \times 16$). We also construct a test matrix $\bm{A}_{test}$ ($1875 \times 14082$), \textit{paper $\times$ words/topics}, from the $2010$ papers, which contains all the papers which needs recommendation. We write ``word/topic" because the content is represented in both ways. \\

Here, $g_{ij}$ is the number of times author $a_i$ has used the word $w_j$ in all of his considered publications. $h_{ij}$ is the number of times word $w_i$ has been used in the conference $c_j$ in total, i.e. considering all the papers that have been accepted in conference $c_j$, all of them combined use the word $w_j$, $h_{ij}$ number of times. We generate the conference matrix by computing the centroid from those entries of the paper matrix which corresponds to this particular conference.\\

\subsubsection{Applied Method}
\label{subsubsec:composition_applied_method}
The algorithm followed is given in the following steps:

\begin{itemize}
	\item We multiply the training matrices, $\bm{A}_{train}$ and $\bm{C}_{train}$, to obtain $\bm{M}_{train}$. The result $\bm{M}_{train}$ is a paper $\times$ conference matrix.
	\item We compute the standardized residual matrix $\bm{S}_{train}$ from $\bm{M}_{train}$ as mentioned in section \ref{subsubsec:implementationCA}.
	\item We then obtain the coordinate matrices (both standard and principal for rows and columns), after decomposing $\bm{S}_{train}$ using SVD.
	\item After this, we multiply the test matrix $\bm{A}_{test}$ with the training matrix $\bm{C}_{train}$ to obtain $\bm{M}_{test}$.
	\item Using the matrix $\bm{M}_{test}$ as a supplementary row matrix, we compute its principal coordinates using the standard column coordinates of $\bm{M}_{train}$.
	\item Then, for each paper in $\bm{M}_{test}$, we compute its similarity with each of the conferences and sort the result.
	\item We then get a ranked list of recommendations for each paper

\end{itemize}

In this method, we multiply the author-words and words-conference matrices and apply CA after that, to recommend a conference to an author. But, this may not capture the relations between the authors and conferences well. An alternative would be to reduce the author-words matrix and the words-conference matrix individually first. Then, defining a transformation from the first subspace to the other might help capture the relations better, which is the next method.

Instead of words, a paper can also be represented in terms of topics. This is more meaningful because if a paper is about information retrieval but does not have much of the IR jargon, then the chances of recommending an IR conference for this paper is less. But, if we capture the topics, then this solves that problem.\\




\subsection{Using Linear Transformation between the reduced-dimensional subspaces}
\label{subsec:linear_trans}
\subsubsection{Data Construction}
\label{subsubsec:linear_trans_data_construction}
The dataset is constructed in the same way as in the previous method. \\

\begin{figure}
\centering
\includegraphics[scale=0.6]{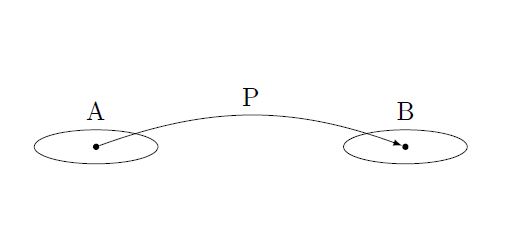}
\caption{Figure showing the mapping $P$ between the two reduced subspaces $A$ and $B$. $A$ is the reduced subspace of author-words matrix and $B$ is the reduced subspace of words-conferences matrix}
\label{subspacemapping}
\end{figure}

The main difference between this method and the previous, however, is that since direct multiplication of the matrices may not capture the relations very well, we reduce each of the matrices author-words and words-conferences to a lower-dimensional subspace and then try to define a transformation $P$ from one to the other i.e. reduced author-words $A$ to the reduced words-conferences $B$. This is illustrated in Figure \ref{subspacemapping}. This, we feel might give a better view of the relations associated between authors and conferences and hence lead to a better recommender system.\\

\subsubsection{Applied Method}
\label{subsubsec:linear_trans_applied_method}
The algorithm followed is given in the following steps:

\begin{itemize}
	\item We compute the principal coordinates for each of the training matrices $\bm{A}_{train}$ and $\bm{C}_{train}$ separately.
	\item We then define a linear transformation from the reduced paper-space to the reduced conference-space as follows: 
	\begin{itemize}
		\item For each paper in $\bm{A}_{train}$, the transformation matrix should map it to the exact conference that it was published in.
		\item So, we collect the principal coordinate vectors of all the training papers in $\bm{F}_{train}$ and then we construct a matrix $\bm{G}_{train}$ corresponding to the principal coordinates of the conferences as follows: the $i^{th}$ row of $\bm{G}_{train}$ is the principal coordinate vector of that conference in which the $i^{th}$ paper in $\bm{F}_{train}$ was published.
		\item Finally the dimensions of $\bm{F}_{train}$ is $3572 \times d_1$ and the dimensions of $\bm{G}_{train}$ is also $3572 \times d_2$, where $d_1$ and $d_2$ are the dimensions of the reduced subspaces, paper-space and conference-space, respectively.
		\item The linear transformation matrix \textbf{T} is then calculated as follows:
				\begin{equation}
					\bm{F}_{train} \bm{T} = \bm{G}_{train} \implies \bm{T} = \bm{F}^{+}_{train} \bm{G}_{train}
				\end{equation}

			where $\bm{F}^{+}_{train}$ is the pseudo-inverse of $\bm{F}_{train}$. The dimensions of the transformation matrix $\bm{T}$ is $d_1 \times d_2$.
		\end{itemize}

	\item Now, using the matrix $\bm{A}_{test}$ as a supplementary row matrix, we compute its principal coordinates using the standard column coordinates of $\bm{A}_{train}$. This matrix has dimensions $1875 \times d_1$
	\item We now need to transform these set of coordinates to the conference space where we can find the similarity easily. This is achieved by multiplying the principal coordinate matrix of the supplementary rows (test papers) with the transformation matrix $\bm{T}$. The dimensions of the resultant matrix is $1875 \times d_2$.
	\item Then, for each paper in the transformed space, we compute the similarity with each of the conferences and sort the result.
	\item We then get a ranked list of recommendations for each paper.

\end{itemize}

This approach is interesting because, to the best of our knowledge, it has never been used in literature of recommender systems. The approach referred to is defining a transformation between the reduced subspaces. In this case, to recommend, we take a paper from its reduced space to the conference space through a linear transformation. \\


\subsection{Content-based Filtering}
\label{subsec:content_based}
\subsubsection{Data Construction}
\label{subsubsec:content_based_data_construction}
The dataset is constructed in the same way as in the previous method. After construction, we require only two matrices, namely, $\bm{A}_{test}$ (paper $\times$ words test matrix) and $\bm{C}_{train}$ (words $\times$ conferences training matrix).\\

\subsubsection{Applied Method}
\label{subsubsec:content_based_applied_method}
This is a very simple method and the algorithm followed is given in the following steps:

\begin{itemize}
	\item For each paper in the test matrix $\bm{A}_{test}$, we compute the similarity with each conference vector in the training matrix $\bm{C}_{train}$. This is done in three ways again as before: euclidean distance, cosine similarity and pearson correlation.
	\item So, for each paper we get a list of similar conferences. We sort them by their similarities and return the result.
	\item We then get a ranked list of recommendations for each paper.
\end{itemize}

This algorithm is a memory-based technique, in contrast to a model-based technique which involves fitting a statistical model over the data and inferring suitable parameters. Although the method is very simple, it gives a very high accuracy compared to the other methods.

\subsection{Collaborative Filtering}
\label{subsec:work_collaborative_filtering}
\subsubsection{Data Construction}
\label{subsubsec:collaborative_filtering_data_construction}
For this method, we construct a paper $\times$ conference matrix, similar to the one in the first method. The difference between this matrix and the one in the first method is that, each row of this matrix represents a paper (multiple authors). We construct two such matrices: one for training and one for testing. Let us call them $\bm{M}_{train}$ and $\bm{M}_{test}$ respectively. For this method, we consider the papers to be the users and the conferences to be the items. So, our objective is to recommend items to the users. The basic idea is to find users similar to the one for whom recommendation is required and then recommend based on what the similar users like.\\

\subsubsection{Applied Method}
\label{subsubsec:collaborative_filtering_applied_method}
The algorithm followed is given in the following steps \cite{segaran2007programming}:

\begin{itemize}
	\item For each paper in $\bm{M}_{test}$, we compute the similar papers from the training matrix $\bm{M}_{train}$. Each paper is a $16$ dimensional vector (representing the frequencies of the conferences) and similarity is calculated using the previous metrics.
	\item We now have to compute the scores for each of the conferences (in other words items). We do this as follows:
		\begin{itemize}
			\item We consider each column of the training matrix $\bm{M}_{train}$. We can think of this representing the ratings given by different users to this particular item.
			\item We then take the ranked list of similar user (papers) to the paper in consideration, and multiply the similarity score of the paper with the corresponding rating of the item.
			\item In more technical terms, we compute the inner product of the similarity vector with the conference column vectors (each of the $16$ in turn).
			\item For the $i^{th}$ conference, say, after computing the inner product, we take sum over all it's entries and then normalize it with the total similarity score (which is the sum of the similarity vector).
			\item The need to do this is because we ignore those users in the inner product who haven't rated the item in question. If this happens, then there is a chance that the items rated by all the users have the maximum score. So, we normalize to make things uniform.
		\end{itemize}		
		      
	\item We now sort the normalized scores for each conference to obtain a ranked list of recommendations.
\end{itemize}

This algorithm is also memory-based technique, in contrast to a model-based technique. This method is commonly used in e-commerce websites for rating-based recommendations.

\subsection{Hybrid Filtering}
\label{subsec:hybrid}
\subsubsection{Data Construction}
\label{subsubsec:hybrid_data_construction}
The data construction for this method is very similar to that in collaborative filtering, with a few additions. We construct the paper $\times$ conference matrix, $\bm{M}_{train}$ as before. In addition, we also construct a paper $\times$ words training matrix, $\bm{A}_{train}$ and paper $\times$ words test matrix, $\bm{A}_{test}$. This method is very similar to that of collaborative filtering algorithm. The only difference is that this method brings in content-analysis too. This is achieved using the paper $\times$ words matrices.

\subsubsection{Applied Method}
\label{subsubsec:hybrid_applied_method}
The algorithm followed is given in the following steps:

\begin{itemize}
	\item For each paper in $\bm{A}_{test}$, we compute the similar papers from the training matrix $\bm{M}_{train}$. Unlike last time, now each paper is a $14082$ dimensional vector (in case of tf-idf) and $400$ dimensional vector (in case of topics). Thus, now, the content of the paper is being brought in as opposed to plain conference-frequency vector. Similarity of each test paper with each of the training papers is calculated using the previous metrics.
	\item The rest of the recommendation process is identical to that in collaborative filtering. This way, we have combined content-based filtering and collaborative filtering.
\end{itemize}

This algorithm is also a memory-based technique, in contrast to a model-based technique.

\section{Evaluation and Results}
\label{sec:eval}
In this section, we detail the evaluation procedures and discuss the results obtained. We have used a total of $7$ metrics to evaluate the performance of the algorithms described above. They were applied on the ranked list of recommendations generated by the above methods:

\begin{itemize}
	\item \textit{Mean Precision at $K$} (MP$@K$): The mean Precision at $K$ for a set of queries is defined as the mean of the Precision at $K$ values for each of those queries. Precision at $K$, $P(K)$, is defined as:
      \begin{equation}
        P(K) = \frac{\text{No. of relevant documents retrieved in the top} \;\; K \;\; \text{results}}{K}
      \end{equation}
     \item \textit{Mean Recall at $K$} (MR$@K$): The mean Recall at $K$ for a set of queries is defined as the mean of the Recall at $K$ values for each of those queries. Recall at $K$, $R(K)$, is defined as:
      \begin{equation}
        R(K) = \frac{\text{No. of relevant documents retrieved in the top} \;\; K \;\; \text{results}}{\text{Total number of relevant documents}}
      \end{equation}
	\item \textit{Mean Average Precision at $K$} (MAP$@K$): Mean average precision at $K$ for a set of queries is the mean of the average precision at $K$ values for each of those queries.
    \begin{equation}
      \text{MAP} = \frac{\sum\limits_{q=1}^Q \text{AveP}(q)}{Q}
    \end{equation}

    where $Q$ is the number of queries. Here $AveP(q)$ is the average precision for the $q^{th}$ query. Average precision is defined as:
      \begin{equation}
        \text{AveP} = \frac{\sum\limits_{k=1}^n \left(P(k) \times rel(k)\right)}{\text{no. of relevant documents}}
      \end{equation}

    where $\text{rel}(k)$ is an indicator function equaling $1$ if the item at rank $k$ is a relevant document, zero otherwise. $P(k)$ is the precision at $k$. 

	\item \textit{Mean Normalized Discounted Cumulative Gain at $P$} (MNDCG$@P$):
    Discounted Cumulative Gain (DCG) at $P$ is defined as:
    \begin{equation}
      \text{DCG}_{P} = \sum\limits_{i=1}^{P} \frac{2^{rel_i}-1}{log_2 (i + 1)}
    \end{equation}

    where $rel_i$ is the relevance score of result $i$. DCG uses a graded scale of relevance and this allows us to have preferences in the predicted results. Let us assume an ideal sequence of predicted results which would yield the maximum $DCG_P$. We call this the ideal $DCG_P$, denoted by $IDCG_P$. The normalized $DCG_P$, $NDCG_P$, is the ratio of the obtained $DCG_P$ with that of the ideal $IDCG_P$. This would thus always yield a value between $0$ and $1$. The mean normalized $DCG_P$ for a set of queries is then the mean of the $NDCG_P$ values for each of those queries.
  \item \textit{Mean Reciprocal Rank} (MRR): The reciprocal rank of a query response is the multiplicative inverse of the rank of the first correct answer. The mean reciprocal rank is the average of the reciprocal ranks of results for a sample of queries $Q$:
  	\begin{equation}
  		\text{MRR} = \frac{1}{\lvert Q \rvert} \sum\limits_{i=1}^{\lvert Q \rvert} \frac{1}{\text{rank}_{i}}
  	\end{equation}
  \item \textit{Mean F-Measure at $K$} (MF-M): The mean F-measure at $K$ for a set of queries is the mean of the F-measures at $K$ for each of those queries. F-measure is defined as the harmonic mean of precision and recall:
  	\begin{equation}
  		F = \frac{2\; . \; \text{precision} \; . \; \text{recall}}{(\text{precision} \; + \; \text{recall})}
  	\end{equation}
  	This is the balanced F-score, where the weights of precision and recall in the harmonic mean are equal. We can also have cases of uneven weights.
  \item \textit{Mean R-Precision} (MR-P):  The mean R-Precision for a set of queries is the mean of the R-Precision values for each of those queries. R-Precision is defined as the Precision at $R$, where $R$ is the number of relevant documents. At this position, the precision and recall values become equal.
\end{itemize}

For the experiments, we have chosen the value of $K$ and $P$ to be $5$. This means that the measures are evaluated (which are $@K$ and $@P$) considering only the top $5$ of the returned results. For the purpose of calculating the metrics, we have defined relevant conferences in two cases:

\begin{enumerate}
	\item A predicted conference is relevant if it is same as the actual conference the paper was originally published in (we have that information from the $2010$ data set). For computing DCG in this case, the relevant conference (which is the original conference) is given a score of $1$ and the rest are given scores $0$.
	\item A predicted conference is relevant if it belongs to the \textit{Special Interest Group (SIG)} of the actual conference the paper was originally published in. For computing DCG in this scenario, the original conference is given a score of $2$, the other conferences in the SIG are given a score of $1$ as they are considered to be partially relevant. The rest of the conferences get a score of $0$.
\end{enumerate}

For calculating similarity to determine the ranking of the retrieved results, we have used three different metrics as previously mentioned:
\begin{itemize}
	\item Euclidean Distance
	\item Cosine Similarity
	\item Pearson Correlation
\end{itemize}

Earlier it was explained that the dimension of the lower-dimensional subspace for an $I \times J$ matrix is $\leq$ min$\left\lbrace I-1,\; J-1\right\rbrace$. Since, we have only $16$ conferences and more than $1000$ papers, the minimum is always $15$. Although the experiments were evaluated for more than one subspace, due to lack of space and vast multitude of results, we only show the results for a $10$-dimensional subspace. We call this $d$. In the case of third method (Linear Transformation), we reduce two matrices independently using CA and hence each can be reduced to a different dimensional subspace. So, for that method, we show the results for $d_1 = 10$, $d_2 = 10, 100$, where $d_1$ is the dimension of the subspace that the Conference x Words/Topics is reduced to and $d_2$ is the dimension of the subspace that the Paper x Words/Topics is reduced to.

The experimental parameters used for LDA are given in Table \ref{lda_param}. For tf-idf, $14082$ words were used.

\begin{table}
\vspace{2ex}
\begin{tabular}{l | l }
\toprule
\textbf{Parameter} & \textbf{Value} \\
\midrule
Number of Iterations &  1000\\\specialrule{\cmidrulewidth}{1pt}{1pt}
Dirichlet Prior $\alpha$ & 0.5\\\specialrule{\cmidrulewidth}{1pt}{1pt}
Number of Topics & 400\\\specialrule{\cmidrulewidth}{1pt}{1pt}
Number of Training Papers & 3572\\\specialrule{\cmidrulewidth}{1pt}{1pt}
Number of Test Papers & 1875\\
\bottomrule
\end{tabular}
\centering
\label{lda_param}
\caption{Experimental Parameters for LDA}
\end{table}

For displaying the results of the experiments, the following conventions are used:
\begin{itemize}
	\item MAP$@5$: \textit{Mean Average Precision at $5$}
	\item MNDCG$@5$: \textit{Mean Normalized Discounted Cumulative Gain at $5$}
	\item MRR: \textit{Mean Reciprocal Rank}
	\item MR-P: \textit{Mean R-Precision}
	\item MF-M: \textit{Mean F-Measure}
	\item MP$@5$: \textit{Mean Precision at $5$}
	\item MR$@5$: \textit{Mean Recall at $5$}
\end{itemize}

Using the above conventions, the evaluations of the experiments are given below:

\subsection{Method 1: Using Author-Conference Matrix}
\label{subsec:method_1}
Here, we present the results for the first method. In this case, evaluation has been conducted with two matrices. The first matrix is the one constructed from the $2010$ test dataset. The second matrix is a null matrix (all entries are 0). The second matrix is required for testing because many authors are common in the training and testing set and it is highly likely that an author, if published in a certain conference, would prefer to publish in it again. Hence, this gives very high accuracy. The only way to really put the method to the test it to consider a new paper, which has not been published in any of the conferences mentioned and then recommend. This is why we considered a null matrix. The results are given below:

\begin{itemize}
	\item \textbf{Case 1}: Using $2010$ test matrix, $d = 10$. The results are displayed in Table \ref{method1_2010}.
\begin{table}
\vspace{2ex}
\begin{tabular}{l | l | l | l | l | l | l}
\toprule
& \multicolumn{2}{c |}{\textbf{Euclid}} & \multicolumn{2}{c |}{\textbf{Cosine}} & \multicolumn{2}{c}{\textbf{Pearson}} \\\cmidrule{2-7}
\textbf{Metrics} & Actual & SIG & Actual & SIG & Actual & SIG \\
\midrule
MAP$@5$ & 0.9483 & 0.6308 & 0.9483 & 0.6308 & 0.9483 & 0.6308 \\\specialrule{\cmidrulewidth}{1pt}{1pt}
MNDCG$@5$ & 0.9613 & 0.8339 & 0.9613 & 0.8339 & 0.9613 & 0.8339\\\specialrule{\cmidrulewidth}{1pt}{1pt}
MRR & 0.9484 & 0.9961 & 0.9484 & 0.9961 & 0.9484 & 0.9961 \\\specialrule{\cmidrulewidth}{1pt}{1pt}
MR-P & 0.9050 & 0.6517 & 0.9050 & 0.6517 & 0.9050 & 0.6517 \\\specialrule{\cmidrulewidth}{1pt}{1pt}
MF-M at $5$ & 0.3328 & 0.5805 & 0.3328 & 0.5805 & 0.3328 & 0.5805 \\\specialrule{\cmidrulewidth}{1pt}{1pt}
MP$@5$ & 0.1997 & 0.5225 & 0.1997 & 0.5225 & 0.1997 & 0.5225\\\specialrule{\cmidrulewidth}{1pt}{1pt}
MR$@5$ & 0.9985 & 0.6531 & 0.9985 & 0.6531 & 0.9985 & 0.6531\\
\bottomrule
\end{tabular}
\centering
\caption{Results for Method 1: Considering the test matrix to be built from \textit{2010} papers, $d = 10$}
\label{method1_2010}
\end{table}

	\item \textbf{Case 2}: Using null test matrix, $d = 10$. The results are displayed in Table \ref{method1_null}.
	\begin{table}
\vspace{2ex}
\begin{tabular}{l | l | l | l | l | l | l}
\toprule
& \multicolumn{2}{c |}{\textbf{Euclid}} & \multicolumn{2}{c |}{\textbf{Cosine}} & \multicolumn{2}{c}{\textbf{Pearson}} \\\cmidrule{2-7}
\textbf{Metrics} & Actual & SIG & Actual & SIG & Actual & SIG \\
\midrule
MAP$@5$ & 0.2072 & 0.2253 & 0.2072 & 0.2253 & 0.2072 & 0.2253 \\\specialrule{\cmidrulewidth}{1pt}{1pt}
MNDCG$@5$ & 0.3042 & 0.3292 & 0.3042 & 0.3292 & 0.3042 & 0.3292\\\specialrule{\cmidrulewidth}{1pt}{1pt}
MRR & 0.2548 & 0.4232 & 0.2548 & 0.4232 & 0.2548 & 0.4232 \\\specialrule{\cmidrulewidth}{1pt}{1pt}
MR-P & 0.0196 & 0.3311 & 0.0196 & 0.3311 & 0.0196 & 0.3311 \\\specialrule{\cmidrulewidth}{1pt}{1pt}
MF-M$@5$ & 0.2013 & 0.3940 & 0.2013 & 0.3940 & 0.2013 & 0.3940 \\\specialrule{\cmidrulewidth}{1pt}{1pt}
MP$@5$ & 0.1208 & 0.3546 & 0.1208 & 0.3546 & 0.1208 & 0.3546\\\specialrule{\cmidrulewidth}{1pt}{1pt}
MR$@5$ & 0.6040 & 0.4433 & 0.6040 & 0.4433 & 0.6040 & 0.4433\\
\bottomrule
\end{tabular}
\centering
\caption{Results for Method 1: Considering the test matrix to be a zero (null) matrix, $d = 10$}
\label{method1_null}
\end{table}
\end{itemize}
As can be seen from the above results, when the input is a null matrix, the method performs poorly.

\subsection{Method 2: Composition of Paper-Words/Topics and Words/Topics-Conference Matrices}
\label{subsec:method_2}
Here we present the results for the second method, which composes two matrices and reduces the dimension. We have two cases: one using tf-idf matrices and one using topic matrices. The results for both are given below:

\begin{itemize}
	\item \textbf{Case 1}: Using tf-idf representation ($14082$ words). $d = 10$. The results are displayed in Table \ref{method2_tfidf}.
\begin{table}
\vspace{2ex}
\begin{tabular}{l | l | l | l | l | l | l}
\toprule
& \multicolumn{2}{c |}{\textbf{Euclid}} & \multicolumn{2}{c |}{\textbf{Cosine}} & \multicolumn{2}{c}{\textbf{Pearson}} \\\cmidrule{2-7}
\textbf{Metrics} & Actual & SIG & Actual & SIG & Actual & SIG \\
\midrule
MAP$@5$ & 0.5800 & 0.7124 & 0.5937 & 0.7778 & 0.5820 & 0.7616 \\\specialrule{\cmidrulewidth}{1pt}{1pt}
MNDCG$@5$ & 0.6573 & 0.7213 & 0.6755 & 0.7571 & 0.6648 & 0.7452\\\specialrule{\cmidrulewidth}{1pt}{1pt}
MRR & 0.5943 & 0.8475 & 0.6041 & 0.8545 & 0.5933 & 0.8507 \\\specialrule{\cmidrulewidth}{1pt}{1pt}
MR-P & 0.3781 & 0.7205 & 0.3829 & 0.7888 & 0.3696 & 0.7686 \\\specialrule{\cmidrulewidth}{1pt}{1pt}
MF-M$@5$ & 0.2956 & 0.6910 & 0.3061 & 0.7477 & 0.3036 & 0.7356 \\\specialrule{\cmidrulewidth}{1pt}{1pt}
MP$@5$ & 0.1773 & 0.6219 & 0.1836 & 0.6729 & 0.1821 & 0.6620\\\specialrule{\cmidrulewidth}{1pt}{1pt}
MR$@R$ & 0.8869 & 0.7774 & 0.9184 & 0.8412 & 0.9109 & 0.8276\\
\bottomrule
\end{tabular}
\centering
\caption{Results for Method 2: Using tf-idf matrices, $d = 10$}
\label{method2_tfidf}
\end{table}

	\item \textbf{Case 2}: Using topic representation ($400$ topics). $d = 10$. The results are displayed in Table \ref{method2_topic}.
	\begin{table}
\vspace{2ex}
\begin{tabular}{l | l | l | l | l | l | l}
\toprule
& \multicolumn{2}{c |}{\textbf{Euclid}} & \multicolumn{2}{c |}{\textbf{Cosine}} & \multicolumn{2}{c}{\textbf{Pearson}} \\\cmidrule{2-7}
\textbf{Metrics} & Actual & SIG & Actual & SIG & Actual & SIG \\
\midrule
MAP$@5$ & 0.3433 & 0.4801 & 0.3880 & 0.5820 & 0.3818 & 0.5715 \\\specialrule{\cmidrulewidth}{1pt}{1pt}
MNDCG$@5$ & 0.4112 & 0.4861 & 0.4600 & 0.5476 & 0.4531 & 0.5383\\\specialrule{\cmidrulewidth}{1pt}{1pt}
MRR & 0.3818 & 0.6330 & 0.4191 & 0.6614 & 0.4136 & 0.6584 \\\specialrule{\cmidrulewidth}{1pt}{1pt}
MR-P & 0.1957 & 0.5068 & 0.2261 & 0.5898 & 0.2218 & 0.5824 \\\specialrule{\cmidrulewidth}{1pt}{1pt}
MF-M$@5$ & 0.2058 & 0.5025 & 0.2259 & 0.5662 & 0.2229 & 0.5534 \\\specialrule{\cmidrulewidth}{1pt}{1pt}
MP$@5$ & 0.1235 & 0.4522 & 0.1355 & 0.5096 & 0.1337 & 0.4981\\\specialrule{\cmidrulewidth}{1pt}{1pt}
MR$@5$ & 0.6176 & 0.5653 & 0.6778 & 0.6370 & 0.6688 & 0.6226\\
\bottomrule
\end{tabular}
\centering
\caption{Results for Method 2: Using topic matrices, $d = 10$}
\label{method2_topic}
\end{table}
\end{itemize}
Here, it is observed that using tf-idf representation for content outperforms its topic counterpart.

\subsection{Method 3: Using Linear Transformation}
\label{subsec:method_3}
Here, we present the results for the third method, which employs a linear transformation between reduced subspaces. We have total of four cases: using tf-idf matrix, topic matrix and different values for $d_1$ and $d_2$. The results for all the cases are given below:

\begin{itemize}
	\item \textbf{Case 1}: Using tf-idf representation ($14082$ words), $d_1 = 10$, $d_2 = 10$. The results are displayed in Table \ref{method3_tfidf_1}.
\begin{table}
\vspace{2ex}
\begin{tabular}{l | l | l | l | l | l | l}
\toprule
& \multicolumn{2}{c |}{\textbf{Euclid}} & \multicolumn{2}{c |}{\textbf{Cosine}} & \multicolumn{2}{c}{\textbf{Pearson}} \\\cmidrule{2-7}
\textbf{Metrics} & Actual & SIG & Actual & SIG & Actual & SIG \\
\midrule
MAP$@5$ & 0.3054 & 0.2566 & 0.4981 & 0.7506 & 0.4933 & 0.6998 \\\specialrule{\cmidrulewidth}{1pt}{1pt}
MNDCG$@5$ & 0.3803 & 0.3920 & 0.5824 & 0.7003 & 0.5687 & 0.6748\\\specialrule{\cmidrulewidth}{1pt}{1pt}
MRR & 0.3520 & 0.5595 & 0.5179 & 0.8318 & 0.5190 & 0.8339 \\\specialrule{\cmidrulewidth}{1pt}{1pt}
MR-P & 0.1882 & 0.3428 & 0.2986 & 0.7178 & 0.3034 & 0.7149 \\\specialrule{\cmidrulewidth}{1pt}{1pt}
MF-M$@5$ & 0.2049 & 0.3984 & 0.2785 & 0.7259 & 0.2643 & 0.6744 \\\specialrule{\cmidrulewidth}{1pt}{1pt}
MP$@5$ & 0.1229 & 0.3586 & 0.1671 & 0.6533 & 0.1586 & 0.6070\\\specialrule{\cmidrulewidth}{1pt}{1pt}
MR$@5$ & 0.6149 & 0.4482 & 0.8357 & 0.8166 & 0.7930 & 0.7588\\
\bottomrule
\end{tabular}
\centering
\caption{Results for Method 3: Using tf-idf matrices, $d_1 = 10$, $d_2 = 10$}
\label{method3_tfidf_1}
\end{table}

	\item \textbf{Case 2}: Using tf-idf representation ($14082$ words), $d_1 = 10$, $d_2 = 100$. The results are displayed in Table \ref{method3_tfidf_2}.
	\begin{table}
\vspace{2ex}
\begin{tabular}{l | l | l | l | l | l | l}
\toprule
& \multicolumn{2}{c |}{\textbf{Euclid}} & \multicolumn{2}{c |}{\textbf{Cosine}} & \multicolumn{2}{c}{\textbf{Pearson}} \\\cmidrule{2-7}
\textbf{Metrics} & Actual & SIG & Actual & SIG & Actual & SIG \\
\midrule
MAP$@5$ & 0.5002 & 0.5347 & 0.5598 & 0.7857 & 0.5520 & 0.7766 \\\specialrule{\cmidrulewidth}{1pt}{1pt}
MNDCG$@5$ & 0.5714 & 0.6117 & 0.6465 & 0.7504 & 0.6419 & 0.7457\\\specialrule{\cmidrulewidth}{1pt}{1pt}
MRR & 0.5225 & 0.8354 & 0.5715 & 0.8917 & 0.5627 & 0.8876 \\\specialrule{\cmidrulewidth}{1pt}{1pt}
MR-P & 0.3354 & 0.5834 & 0.3397 & 0.7669 & 0.5627 & 0.8876 \\\specialrule{\cmidrulewidth}{1pt}{1pt}
MF-M$@5$ & 0.2618 & 0.5442 & 0.3013 & 0.7454 & 0.3032 & 0.7461 \\\specialrule{\cmidrulewidth}{1pt}{1pt}
MP$@5$ & 0.1571 & 0.4898 & 0.1808 & 0.6709 & 0.1819 & 0.6715\\\specialrule{\cmidrulewidth}{1pt}{1pt}
MR$@5$ & 0.7856 & 0.6122 & 0.9040 & 0.8386 & 0.9098 & 0.8394\\
\bottomrule
\end{tabular}
\centering
\caption{Results for Method 3: Using tf-idf matrices, $d_1 = 10$, $d_2 = 100$}
\label{method3_tfidf_2}
\end{table}

\item \textbf{Case 3}: Using topic representation ($400$ topics), $d_1 = 10$, $d_2 = 10$. The results are displayed in Table \ref{method3_topic_1}.
\begin{table}
\vspace{2ex}
\begin{tabular}{l | l | l | l | l | l | l}
\toprule
& \multicolumn{2}{c |}{\textbf{Euclid}} & \multicolumn{2}{c |}{\textbf{Cosine}} & \multicolumn{2}{c}{\textbf{Pearson}} \\\cmidrule{2-7}
\textbf{Metrics} & Actual & SIG & Actual & SIG & Actual & SIG \\
\midrule
MAP$@5$ & 0.4087 & 0.5203 & 0.4265 & 0.6133 & 0.4262 & 0.6063 \\\specialrule{\cmidrulewidth}{1pt}{1pt}
MNDCG$@5$ & 0.4708 & 0.5347 & 0.4932 & 0.5779 & 0.4934 & 0.5771\\\specialrule{\cmidrulewidth}{1pt}{1pt}
MRR & 0.4416 & 0.6827 & 0.4548 & 0.6914 & 0.4543 & 0.6945 \\\specialrule{\cmidrulewidth}{1pt}{1pt}
MR-P & 0.2522 & 0.5326 & 0.2565 & 0.6178 & 0.2581 & 0.6097 \\\specialrule{\cmidrulewidth}{1pt}{1pt}
MF-M$@5$ & 0.2188 & 0.5245 & 0.2305 & 0.5870 & 0.2314 & 0.5870 \\\specialrule{\cmidrulewidth}{1pt}{1pt}
MP$@5$ & 0.1313 & 0.4721 & 0.1383 & 0.5283 & 0.1388 & 0.5283\\\specialrule{\cmidrulewidth}{1pt}{1pt}
MR$@5$ & 0.6565 & 0.5901 & 0.6917 & 0.6604 & 0.6944 & 0.6604\\
\bottomrule
\end{tabular}
\centering
\caption{Results for Method 3: Using topic matrices, $d_1 = 10$, $d_2 = 10$}
\label{method3_topic_1}
\end{table}

	\item \textbf{Case 4}: Using topic representation ($400$ topics), $d_1 = 10$, $d_2 = 100$. The results are displayed in Table \ref{method3_topic_2}.
	\begin{table}
\vspace{2ex}
\begin{tabular}{l | l | l | l | l | l | l}
\toprule
& \multicolumn{2}{c |}{\textbf{Euclid}} & \multicolumn{2}{c |}{\textbf{Cosine}} & \multicolumn{2}{c}{\textbf{Pearson}} \\\cmidrule{2-7}
\textbf{Metrics} & Actual & SIG & Actual & SIG & Actual & SIG \\
\midrule
MAP$@5$ & 0.4525 & 0.5531 & 0.4614 & 0.6325 & 0.4566 & 0.6219 \\\specialrule{\cmidrulewidth}{1pt}{1pt}
MNDCG$@5$ & 0.5152 & 0.5724 & 0.5259 & 0.6038 & 0.5210 & 0.5988\\\specialrule{\cmidrulewidth}{1pt}{1pt}
MRR & 0.4816 & 0.7179 & 0.4871 & 0.7156 & 0.4836 & 0.7192 \\\specialrule{\cmidrulewidth}{1pt}{1pt}
MR-P & 0.2965 & 0.5596 & 0.2970 & 0.6356 & 0.2938 & 0.6201 \\\specialrule{\cmidrulewidth}{1pt}{1pt}
MF-M$@5$ & 0.2343 & 0.5511 & 0.2394 & 0.6021 & 0.2376 & 0.5985 \\\specialrule{\cmidrulewidth}{1pt}{1pt}
MP$@5$ & 0.1405 & 0.4960 & 0.1436 & 0.5419 & 0.1426 & 0.5386\\\specialrule{\cmidrulewidth}{1pt}{1pt}
MR$@5$ & 0.7029 & 0.6200 & 0.7184 & 0.6774 & 0.7130 & 0.6733\\
\bottomrule
\end{tabular}
\centering
\caption{Results for Method 3: Using topic matrices, $d_1 = 10$, $d_2 = 100$}
\label{method3_topic_2}
\end{table}
\end{itemize}

As can be observed, just like the third method, the tf-idf representation overall outperforms the topic representation. In some cases, the topic representation outperforms its counterpart, for example when considering the euclidean distance metric, and $d_1 = 10$, $d_2 = 10$.

\subsection{Method 4: Content-based Filtering}
\label{subsec:method_4}
Here, we present the results for content-based filtering. This method does not use any dimensionality reduction techniques and is a \textit{memory-based} (which is different from \textit{model-based} methods where we try to fit a statistical model to the data and infer the parameters, which are then used to determine the results) method. There are two cases here too: using tf-idf and topic representations.

\begin{itemize}
	\item \textbf{Case 1}: Using tf-idf representation ($14082$ words). The results are displayed in Table \ref{method4_tfidf}.
\begin{table}
\vspace{2ex}
\begin{tabular}{l | l | l | l | l | l | l}
\toprule
& \multicolumn{2}{c |}{\textbf{Euclid}} & \multicolumn{2}{c |}{\textbf{Cosine}} & \multicolumn{2}{c}{\textbf{Pearson}} \\\cmidrule{2-7}
\textbf{Metrics} & Actual & SIG & Actual & SIG & Actual & SIG \\
\midrule
MAP$@5$ & 0.6477 & 0.7530 & 0.6637 & 0.7758 & 0.6622 & 0.7762 \\\specialrule{\cmidrulewidth}{1pt}{1pt}
MNDCG$@5$ & 0.7200 & 0.7718 & 0.7380 & 0.7920 & 0.7367 & 0.7916\\\specialrule{\cmidrulewidth}{1pt}{1pt}
MRR & 0.6557 & 0.9034 & 0.6695 & 0.9166 & 0.6683 & 0.9156 \\\specialrule{\cmidrulewidth}{1pt}{1pt}
MR-P & 0.4544 & 0.7516 & 0.4613 & 0.7790 & 0.4597 & 0.7797 \\\specialrule{\cmidrulewidth}{1pt}{1pt}
MF-M$@5$ & 0.3112 & 0.7191 & 0.3191 & 0.7413 & 0.3187 & 0.7421 \\\specialrule{\cmidrulewidth}{1pt}{1pt}
MP$@5$ & 0.1867 & 0.6472 & 0.1914 & 0.6672 & 0.1912 & 0.6679\\\specialrule{\cmidrulewidth}{1pt}{1pt}
MR$@5$ & 0.9338 & 0.8090 & 0.9573 & 0.8340 & 0.9562 & 0.8349\\
\bottomrule
\end{tabular}
\centering
\caption{Results for Method 4: Using tf-idf matrices}
\label{method4_tfidf}
\end{table}

	\item \textbf{Case 2}: Using topic representation ($400$ topics). The results are displayed in Table \ref{method4_topic}.
	\begin{table}
\vspace{2ex}
\begin{tabular}{l | l | l | l | l | l | l}
\toprule
& \multicolumn{2}{c |}{\textbf{Euclid}} & \multicolumn{2}{c |}{\textbf{Cosine}} & \multicolumn{2}{c}{\textbf{Pearson}} \\\cmidrule{2-7}
\textbf{Metrics} & Actual & SIG & Actual & SIG & Actual & SIG \\
\midrule
MAP$@5$ & 0.4371 & 0.5913 & 0.4385 & 0.5947 & 0.4215 & 0.5955 \\\specialrule{\cmidrulewidth}{1pt}{1pt}
MNDCG$@5$& 0.5138 & 0.5879 & 0.5157 & 0.5900 & 0.4975 & 0.5779\\\specialrule{\cmidrulewidth}{1pt}{1pt}
MRR & 0.4650 & 0.7222 & 0.4657 & 0.7206 & 0.4494 & 0.6973 \\\specialrule{\cmidrulewidth}{1pt}{1pt}
MR-P & 0.2586 & 0.6042 & 0.2592 & 0.6073 & 0.2464 & 0.6077 \\\specialrule{\cmidrulewidth}{1pt}{1pt}
MF-M$@5$ & 0.2481 & 0.5880 & 0.2494 & 0.5916 & 0.2421 & 0.5876 \\\specialrule{\cmidrulewidth}{1pt}{1pt}
MP$@5$ & 0.1489 & 0.5292 & 0.1496 & 0.5324 & 0.1452 & 0.5288\\\specialrule{\cmidrulewidth}{1pt}{1pt}
MR$@5$ & 0.7445 & 0.6616 & 0.7482 & 0.6656 & 0.7264 & 0.6610\\
\bottomrule
\end{tabular}
\centering
\caption{Results for Method 4: Using topic matrices}
\label{method4_topic}
\end{table}
\end{itemize}

As we can see from the above results, tf-idf representation again outperforms the topic representation.

\subsection{Method 5: Collaborative Filtering}
\label{subsec:method_5}
Here, we present the results for collaborative filtering. This method does not use any dimensionality reduction techniques and is \textit{memory-based}, just like content-based filtering. This method does not use content, rather just uses similarity measures to determine the recommendations. The results are displayed in Table \ref{method5}.

\begin{table}
\vspace{2ex}
\begin{tabular}{l | l | l | l | l | l | l}
\toprule
& \multicolumn{2}{c |}{\textbf{Euclid}} & \multicolumn{2}{c |}{\textbf{Cosine}} & \multicolumn{2}{c}{\textbf{Pearson}} \\\cmidrule{2-7}
\textbf{Metrics} & Actual & SIG & Actual & SIG & Actual & SIG \\
\midrule
MAP$@5$ & 0.1224 & 0.1531 & 0.0573 & 0.1066 & 0.1039 & 0.1161 \\\specialrule{\cmidrulewidth}{1pt}{1pt}
MNDCG$@5$ & 0.1802 & 0.2144 & 0.0740 & 0.1371 & 0.1519 & 0.1861\\\specialrule{\cmidrulewidth}{1pt}{1pt}
MRR & 0.1945 & 0.3131 & 0.1456 & 0.3862 & 0.1829 & 0.3089 \\\specialrule{\cmidrulewidth}{1pt}{1pt}
MR-P & 0.0213 & 0.2696 & 0.0192 & 0.1773 & 0.0234 & 0.1544 \\\specialrule{\cmidrulewidth}{1pt}{1pt}
MF-M$@5$ & 0.1192 & 0.2692 & 0.0414 & 0.1765 & 0.1004 & 0.2435 \\\specialrule{\cmidrulewidth}{1pt}{1pt}
MP$@5$ & 0.0715 & 0.2423 & 0.0248 & 0.1589 & 0.0602 & 0.2192\\\specialrule{\cmidrulewidth}{1pt}{1pt}
MR$@5$ & 0.3578 & 0.3029 & 0.1242 & 0.1986 & 0.3013 & 0.2740\\
\bottomrule
\end{tabular}
\centering
\caption{Results for Method 5: Collaborative Filtering}
\label{method5}
\end{table}

\subsection{Method 6: Hybrid Filtering}
\label{subsec:method_6}
Here, we present the results for hybrid filtering, which is a hybrid of content-based filtering and collaborative filtering. This method does not use any dimensionality reduction techniques and is also \textit{memory-based}. This method combines the good qualities of both content-based filtering and collaborative filtering. The results are displayed in Table \ref{method6_tfidf}.

\begin{itemize}
	\item \textbf{Case 1}: Using tf-idf representation ($14082$ words). The results are displayed in Table \ref{method6_tfidf}.
\begin{table}
\vspace{2ex}
\begin{tabular}{l | l | l | l | l | l | l}
\toprule
& \multicolumn{2}{c |}{\textbf{Euclid}} & \multicolumn{2}{c |}{\textbf{Cosine}} & \multicolumn{2}{c}{\textbf{Pearson}} \\\cmidrule{2-7}
\textbf{Metrics} & Actual & SIG & Actual & SIG & Actual & SIG \\
\midrule
MAP$@5$ & 0.0566 & 0.1083 & 0.1036 & 0.1190 & 0.1037 & 0.1191 \\\specialrule{\cmidrulewidth}{1pt}{1pt}
MNDCG$@5$ & 0.0734 & 0.1380 & 0.1497 & 0.1860 & 0.1499 & 0.1872\\\specialrule{\cmidrulewidth}{1pt}{1pt}
MRR & 0.1449 & 0.3930 & 0.1867 & 0.3223 & 0.1865 & 0.3231 \\\specialrule{\cmidrulewidth}{1pt}{1pt}
MR-P & 0.0192 & 0.1773 & 0.0213 & 0.1817 & 0.0213 & 0.1813 \\\specialrule{\cmidrulewidth}{1pt}{1pt}
MF-M$@5$ & 0.0414 & 0.1765 & 0.0972 & 0.2388 & 0.0974 & 0.2417 \\\specialrule{\cmidrulewidth}{1pt}{1pt}
MP$@5$ & 0.0248 & 0.1589 & 0.0583 & 0.2149 & 0.0584 & 0.2176\\\specialrule{\cmidrulewidth}{1pt}{1pt}
MR$@5$ & 0.1242 & 0.1986 & 0.2917 & 0.2686 & 0.2922 & 0.2720\\
\bottomrule
\end{tabular}
\centering
\caption{Results for Method 6: Using tf-idf matrices}
\label{method6_tfidf}
\end{table}

	\item \textbf{Case 2}: Using topic representation ($400$ topics). The results are displayed in Table \ref{method6_topic}.
	\begin{table}
\vspace{2ex}
\begin{tabular}{l | l | l | l | l | l | l}
\toprule
& \multicolumn{2}{c |}{\textbf{Euclid}} & \multicolumn{2}{c |}{\textbf{Cosine}} & \multicolumn{2}{c}{\textbf{Pearson}} \\\cmidrule{2-7}
\textbf{Metrics} & Actual & SIG & Actual & SIG & Actual & SIG \\
\midrule
MAP$@5$ & 0.0519 & 0.0984 & 0.1241 & 0.1577 & 0.1006 & 0.1486 \\\specialrule{\cmidrulewidth}{1pt}{1pt}
MNDCG$@5$ & 0.0720 & 0.1270 & 0.1892 & 0.2388 & 0.1477 & 0.2032\\\specialrule{\cmidrulewidth}{1pt}{1pt}
MRR & 0.1334 & 0.3109 & 0.1945 & 0.3237 & 0.1813 & 0.3097 \\\specialrule{\cmidrulewidth}{1pt}{1pt}
MR-P & 0.0192 & 0.1773 & 0.0218 & 0.2688 & 0.0245 & 0.2362 \\\specialrule{\cmidrulewidth}{1pt}{1pt}
MF-M$@5$ & 0.0449 & 0.1765 & 0.1308 & 0.3229 & 0.0983 & 0.2798 \\\specialrule{\cmidrulewidth}{1pt}{1pt}
MP$@5$ & 0.0269 & 0.1589 & 0.0785 & 0.2906 & 0.0589 & 0.2518\\\specialrule{\cmidrulewidth}{1pt}{1pt}
MR$@5$ & 0.1349 & 0.1986 & 0.3925 & 0.3633 & 0.2949 & 0.3148\\
\bottomrule
\end{tabular}
\centering
\caption{Results for Method 6: Using topic matrices}
\label{method6_topic}
\end{table}
\end{itemize}

Here also, we observe that the tf-idf representation outperforms its topic counterpart.

\section{Conclusions and Future Work}
\label{sec:conclusions_future}
\subsection{Conclusions}
\label{subsec:conclusions}
Although each of the above methods has its own merits, from the results obtained we observe the following:
\begin{itemize}
  \item The content-based methods proposed easily beat popular methods like collaborative filtering. This shows that for this system, considering content is vital. Computing similarities with content in hybrid filtering also did not prove to be very helpful, as the remainder of the procedure is identical to collaborative filtering.
	\item The first method, involving just the conference-frequencies and not the content, is seen to perform poorly when it comes to recommending for new authors. The very high accuracy when considering the $2010$ test matrix can be attributed to the fact that authors tend to publish in conferences where they have published before. Our proposed content-based methods involving CA work equally well with old/new authors because only the content of the paper is taken into consideration and not their prior publication counts in the conferences.
	\item Content-based filtering is seen to outperform the CA-based methods. This may be attributed to the fact that there is a certain amount of information loss during the dimensionality reduction phase, while content-based filtering utilizes the ``pure" raw content.
	\item In the results obtained, using tf-idf for content proved to be better than using topics. This may be due to considering a much larger number of words in tf-idf representation ($14082$) than it's topic counterpart ($400$). Also, the method of generating the topic matrices may have influenced the results.
	\item Lastly, we observe that cosine similarity proves to be the best measure to calculate the similarities.
\end{itemize}

\subsection{Future Work}
\label{subsec:future }
We can improve accuracy of the content-analysis techniques by considering more attributes for the papers such as keywords that can very well help in the recommendations. Improvements might be seen, if we can incorporate the network information of the authors along with the content of the paper into the recommender system. The citation information of the papers can also serve as a good feature. 

\newpage

\bibliography{bib}

\end{document}